\documentclass[final,5p,times,twocolumn]{elsarticle}

\usepackage{lineno}
\usepackage{hyperref}
\usepackage{amsmath,amssymb}
\usepackage{graphicx}
\usepackage{booktabs}
\usepackage{array}
\usepackage{tabularx}
\usepackage{multirow}
\usepackage{xcolor}
\usepackage{url}
\usepackage{microtype}
\usepackage[capitalise,noabbrev]{cleveref}
\usepackage{subcaption}
\usepackage{algorithm}
\usepackage{threeparttable}
\usepackage{algpseudocode}
\usepackage{fancyhdr}

\modulolinenumbers[5]

\hypersetup{
  colorlinks=true,
  linkcolor=blue!70!black,
  citecolor=blue!70!black,
  urlcolor=blue!70!black,
  pdfauthor={Authors},
  pdftitle={BRIDGE and TCH-Net: Benchmark Reference for IoT Domain Generalisation Evaluation and Cross-Branch Gated Attention for Botnet Detection},
}

\newcolumntype{C}[1]{>{\centering\arraybackslash}p{#1}}
\newcolumntype{L}[1]{>{\raggedright\arraybackslash}p{#1}}

\newcommand{\mF}{\mathrm{F1}}
\newcommand{\mA}{\mathrm{AUC}}
\newcommand{\mM}{\mathrm{MCC}}

\newcommand{\cbgaf}{\textsc{CB-GAF}}

\newcommand{\tchnet}{TCH-Net}
\newcommand{\bridge}{\textsc{BRIDGE}}
\newcommand{\plusminus}{\pm}

\journal{Journal of Network and Computer Applications}

\begin{document}

%% Force fancy style on first page too
\thispagestyle{fancy}

\begin{frontmatter}

\title{BRIDGE and TCH-Net: Heterogeneous Benchmark and Multi-Branch Baseline for Cross-Domain IoT Botnet Detection}

\author[kiit]{Ammar Bhilwarawala\corref{cor1}}
\ead{ammarbhilwarawala@gmail.com}

\author[kiit]{Likhamba Rongmei}
\ead{23052578@kiit.ac.in}

\author[kiit]{Harsh Sharma}
\ead{23052573@kiit.ac.in}

\author[kiit]{Arya Jena}
\ead{23053346@kiit.ac.in}

\author[kiit]{Kaushal Singh}
\ead{23052575@kiit.ac.in}

\author[kiit]{Jayashree Piri}
\ead{jayashree.pirifcs@kiit.ac.in}

\author[kiit]{Raghunath Dey}
\ead{raghunath.deyfcs@kiit.ac.in}

\cortext[cor1]{Corresponding author.}

\address[kiit]{School of Computer Engineering,
Kalinga Institute of Industrial Technology~(KIIT),
Bhubaneswar, Odisha~751024, India}

%% Abstract + Keywords

\begin{abstract}

Despite years of progress in IoT botnet detection, the field has been quietly building on a shaky foundation: the overwhelming majority of published systems are evaluated on a single dataset, producing performance estimates that simply do not hold when the network environment changes. Compounding this, the heterogeneous feature spaces of available IoT security datasets have made principled multi-dataset training practically impossible without either discarding semantic interpretability or introducing silent data integrity violations. No prior work has addressed both problems together with a formally specified, reproducible methodology.
This paper makes two primary contributions toward fixing that. First, we introduce BRIDGE (Benchmark Reference for IoT Domain Generalisation Evaluation), the first formally specified heterogeneous multi-dataset benchmark for IoT intrusion detection, unifying five structurally distinct publicly available datasets, CICIDS-2017, CIC-IoT-2023, Bot-IoT, Edge-IIoTset, and N-BaIoT, through a 46-feature semantic canonical vocabulary grounded in CICFlowMeter nomenclature, with genuine-equivalence-only feature mapping, explicit zero-filling for absent features, and full per-dataset coverage disclosure spanning 15\% to 93\%. A leave-one-dataset-out (LODO) evaluation protocol reveals, for the first time with a formally reproducible methodology, just how large the generalisation gap really is: all five evaluated deep learning architectures achieve mean LODO F1 in the range 0.39–0.47, and we establish the first formally quantified community generalisation baseline at mean LODO F1 = 0.5577, which is a finding that we believe will reframe the research agenda from single-benchmark optimisation toward cross-environment generalisation and domain adaptation.
Second, we propose TCH-Net as a strong and well-characterised baseline for BRIDGE: a multi-branch neural architecture integrating a three-path Temporal branch with residual convolutional-BiGRU, stride-downsampled BiGRU, and full-resolution pre-LayerNorm Transformer encoders for multi-scale attack pattern capture, a provenance-conditioned Contextual branch, and an aggregate Statistical branch, fused via a novel Cross-Branch Gated Attention Fusion (CB-GAF) mechanism with learnable per-branch sigmoid gates that enable dynamic, feature-wise cross-branch information mixing. Evaluated across five independent random seeds on BRIDGE, TCH-Net achieves F1 = 0.8296 ± 0.0028, AUC = 0.9380 ± 0.0025, and MCC = 0.6972 ± 0.0056, outperforming all twelve baseline models with statistical significance (p < 0.05, paired Wilcoxon signed-rank test) and attaining the highest cross-dataset LODO F1 among all evaluated architectures. BRIDGE, its canonical vocabulary specification, and the complete experimental pipeline are publicly released at \url{https://github.com/Ammar-ss/TCH-Net} to facilitate reproducible community evaluation and progress on the cross-dataset generalisation challenge that BRIDGE makes, for the first time, precisely measurable.

\end{abstract}

 
\end{frontmatter}

\section{Introduction}
\label{sec:intro}
 
The Internet of Things~(IoT) has transformed how physical devices interact with digital  infrastructure, extending connectivity from industrial sensors and smart-home appliances to medical monitors and autonomous vehicles~\cite{grammatikis2018}. This expansion is accompanied by a commensurate growth in the cyber attack surface: IoT devices are typically resource-constrained, ship with minimal security hardening, and operate across heterogeneous communication protocols that resist uniform monitoring, making them an attractive target for large-scale botnet conscription.
 
Botnets represent one of the most operationally damaging threat classes in the contemporary threat landscape.
The Mirai botnet~\cite{kolias2017}, first observed in 2016, demonstrated that hundreds of thousands of misconfigured IoT devices could be marshalled into a coordinated distributed denial-of-service~(DDoS) platform capable of generating traffic volumes exceeding 600~Gbps, sufficient to disrupt a major chunk of internet infrastructure for extended periods of time \cite{antonakakis2017}. Subsequent variants: Satori, Okiru, and Masuta confirm that the Mirai template is iteratively refined to exploit newly discovered device classes and vulnerability surfaces. Beyond DDoS amplification, modern IoT botnets serve as infrastructure for credential stuffing, crypto-mining, spam relay, and lateral movement within enterprise and industrial networks.
The economic cost of IoT-facilitated cyberattacks is estimated to be in hundreds of billions of dollars annually~\cite{anderson2019}, and results in the disruption of critical infrastructure which elevates these threats beyond financial harm to the matter of public security and privacy.
 
\subsection{Limitations of Conventional Detection Approaches}
\label{subsec:intro_conv}
 
The traditional network-level defence against these botnet activities is the Intrusion Detection System~(IDS), which analyzes the traffic to distinguish malicious from benign interactions.
Signature-based IDS platforms such as Snort~\cite{roesch1999} maintains a curated rule databases matching known attack patterns. While highly precise for cataloged threats, they are still unable to consistently detect zero-day exploits, polymorphic malware, or previously unseen botnet C\&C protocols. Anomaly-based detection~\cite{garcia2014} circumvents the zero-day blind spots but suffers from elevated false positive rates in IoT environments where diverse device behaviour renders any single baseline inadequate~\cite{meidan2018}. Classical machine learning models achieve strong benchmark performance, yet treat flows as independent samples, discarding temporal ordering, the very structure in which coordinated attack behaviour is encoded~\cite{sommer2010}.
 
\subsection{The Single-Dataset Evaluation Crisis}
\label{subsec:intro_single}
 
Recurrent neural architectures~\cite{hochreiter1997,cho2014}, convolutional networks~\cite{lecun1998}, and transformer-based models~\cite{vaswani2017} have enabled IDS  to exploit sequential dependencies that flat feature vectors cannot capture. Despite this progress, the vast majority of published systems are evaluated on a \emph{single} benchmark dataset, producing optimistic estimates tuned to one capture environment, one time period, and one attack toolkit.
Ring et al.~\cite{ring2019} surveyed 34 network IDS datasets and found that single-dataset evaluation is the dominant paradigm, with feature naming inconsistencies, labelling methodology differences, and capture tool variations identified as primary obstacles to principled multi-dataset comparison. Compounding this, widely-used benchmarks including CICIDS-2017~\cite{sharafaldin2018} contain systematic labelling artifact and CICFlowMeter implementation errors that artificially inflate reported metrics~\cite{engelen2021}. As Sommer and Paxson~\cite{sommer2010} demonstrated empirically, models trained in a closed-world benchmark exhibit dramatic performance degradation outside it,
a limitation particularly acute in heterogeneous IoT environments where device populations and attack toolkits shift continuously. The field therefore lacks a reliable answer to a fundamental question: \emph{how well do IoT botnet detection systems actually  generalize across the diverse network environments in which they must operate?}
 
\subsection{The Feature Heterogeneity Problem}
\label{subsec:intro_hetero}
 
A natural response to single-dataset fragility is multi-dataset training, but the network security dataset ecosystem is characterized by profound feature-space heterogeneity: CICFlowMeter datasets~\cite{lashkari2017} export bidirectional flow statistics, Argus~\cite{koroniotis2018} produces session-level records, Wireshark~\cite{ferrag2022} captures packet-level attributes, and Kitsune~\cite{meidan2018} produces statistical fingerprint vectors with no flow-level correspondence. Existing multi-dataset approaches either apply PCA~\cite{jolliffe2002}, discarding semantic interpretability, or employ ad-hoc proxy mappings that introduce silent data integrity violations. Neither of which provides a principled, reproducible solution, which is why single-dataset evaluation still persists.
 
\subsection{Contributions of This Work}
\label{subsec:intro_contributions}
 
This paper addresses both the evaluation crisis and the feature heterogeneity problem through two interconnected contributions:
 
\begin{enumerate}
 
\item \textbf{\bridge{}: A Benchmark Reference for IoT Domain Generalisation Evaluation.}
We introduce \bridge{}, the first formally specified evaluation benchmark unifying five
structurally distinct IoT network security datasets through a principled
feature alignment framework.
\bridge{} comprises:
\begin{itemize}
    \item A \textbf{46-feature semantic canonical vocabulary} grounded in CICFlowMeter nomenclature, with genuine-equivalence-only mapping constraints, explicit zero-filling for absent features, and full per-dataset coverage disclosure spanning 15\% to 93\% across the five datasets.
    \item A \textbf{reproducible preprocessing pipeline} including class balancing, shared RbustScaler normalisation, sliding-window sequence construction, and verified leakage-free train/test splitting.
    \item A \textbf{leave-one-dataset-out~(LODO) evaluation protocol} providing the first formally quantified cross-dataset generalisation benchmark in heterogeneous IoT intrusion detection, establishing mean LODO~F1~$= 0.5577$ as a rigorous \bridge{} community baseline that exposes the domain shift as a primary challenge in this sector.
\end{itemize}
 
\item \textbf{TCH-Net: A Multi-Branch Architecture for Multi-Dataset Botnet
Detection.}
We propose TCH-Net as a deep neural architecture comprising of three specialized parallel branches designed to exploit distinct modalities of network flow  nformation:
\begin{itemize}
    \item \textbf{Cross-Branch Gated Attention Fusion~(CB-GAF)}: a novel fusion mechanism in which each branch queries the remaining two via scaled dot-product attention modulated by learnable per-branch sigmoid gates, enabling dynamic and asymmetric cross-branch information mixing. Component ablation using \textsc{TCHNovAbl} confirms CB-GAF is necessary; its removal degrades F1 by ${\sim}0.054$ relative to the full model (inclusive of proxy architectural gap).
    \item \textbf {Three-Path Temporal Encoding: a T-branch consisting of three parallel encoders, (i) a residual depthwise-separable convolutional stack with Squeeze-Excitation recalibration followed by a two-layer BiGRU capturing both local and medium-range sequential patterns; (ii) a stride-downsampled convolutional projection followed by a single-layer BiGRU capturing coarse-scale dynamics; and (iii) a full-resolution two-layer pre-LayerNorm Transformer encoder with CLS-token classification capturing global temporal context, fused via multi-head self-attention across a shared 8-step temporal grid into $\mathbf{h}^T \in \mathbb{R}^{512}$.}
    Component ablation confirms MSTE is necessary; its removal
degrades F1 by ${\sim}0.054$ relative to the full model
(inclusive of proxy architectural gap).

    \item \textbf{Dual Domain Embedding}: a contextual branch encoding dataset identity and device category as learned dense embeddings, conditioning CB-GAF's fusion behaviour on input provenance and enabling the model to calibrate cross-branch information mixing based on the feature coverage
    profile of each source dataset.
\end{itemize}
After evaluated over all five independent random seeds on \bridge{}, TCH-Net achieves F1~$= 0.8296 \pm 0.0028$, AUC~$= 0.9380 \pm 0.0025$, and MCC~$= 0.6972 \pm 0.0056$, outperforming all twelve previously evaluated baseline models with statistical significance ($p < 0.05$, paired Wilcoxon signed-rank test).
 
\end{enumerate}
 
\textbf{Paper organisation.} Section~\ref{sec:related} surveys related work. Section~\ref{sec:datasets} details \bridge{} and its preprocessing pipeline. Section~\ref{sec:architecture} presents the TCH-Net architecture. Section~\ref{sec:results} reports all experimental results. Section~\ref{sec:discussion} discusses findings and limitations. Section~\ref{sec:conclusion} concludes.

\section{Related Work}
\label{sec:related}

\subsection{Classical Machine Learning for Network IDS}
\label{subsec:rel_classical}

Decision tree ensembles, especially random forests~\cite{breiman2001}, became a dominant paradigm for flow-based IDS owing to their resistance to irrelevant features, native handling of mixed-type inputs, and interpretable feature importance scores. Random forest classifiers trained on CICFlowMeter-derived features routinely achieve high accuracy on single-dataset benchmarks, largely because many exported features contain substantial discriminative redundancy that tree ensembles exploit efficiently. Gradient-boosted decision trees, epitomised by XGBoost~\cite{chen2016xgboost},
extend this through sequential residual fitting and have been applied to both binary and multi-class intrusion detection with similar results~\cite{ring2019}.

These models treat flows as an independent set of samples and rely on handcrafted features, both of which limits generalization when device population or attack toolkit shifts~\cite{sommer2010}.

\subsection{Recurrent and Convolutional Deep Learning for IDS}
\label{subsec:rel_deep}

LSTM networks~\cite{hochreiter1997} and their bidirectional~\cite{imrana2021} and GRU~\cite{cho2014} variants model temporal dependencies across flow sequences. Applied to CICFlowMeter data, they report F1 of 0.97 to 0.99 on CICIDS-2017~\cite{imrana2021}, performance that does not transfer to held-out environments, as our LODO results confirmed.

One-dimensional CNNs~\cite{lecun1998} extract local temporal motifs; CNN-LSTM hybrids extend this with a long-range memory. A consistent limitation of all single-path architectures is the absence of principled mechanisms for fusing temporal, statistical, and provenance modalities of network flows, which TCH-Net addresses through its three-branch CB-GAF design.

\subsection{Transformer-Based Approaches}
\label{subsec:rel_transformer}

The transformer architecture~\cite{vaswani2017}, built on scaled multi-head self-attention, theoretically allows the attending of arbitrarily distant positions within a sequence window without the sequential bottleneck of recurrent computation. Recent works have applied transformer encoders to flow-level network traffic classification, consistently finding that self-attention provides modest
improvements over BiLSTM baselines on single-dataset benchmarks when model capacity is held constant~\cite{akuthota2025}. A critical challenge for transformer-based IDS is data volume: transformer models are notoriously data-hungry, and the effective training sets available after
class-balancing can leave transformers under-trained relative to their capacity, leading to elevated variance across random seeds. In TCH-Net, the Transformer is deliberately restricted to a fixed 32-step window within the T-branch, maintaining a favourable data-to-parameter ratio while contributing global temporal context alongside the recurrent paths.

\subsection{IoT-Specific Intrusion Detection Systems}
\label{subsec:rel_iot}

N-BaIoT~\cite{meidan2018} pioneered deep autoencoders for device-level IoT botnet detection, representing each device's traffic as a high-dimensional vector of sliding-window statistical features and training a per-device auto-encoder on benign traffic to detect botnet-induced reconstruction anomalies. N-BaIoT achieves near-perfect detection on trained device types but requires
per-device model training and cannot be generalized to unseen device classes.

Kitsune~\cite{mirsky2018} trains an ensemble of feature-group auto-encoders incrementally on streaming traffic, though concept drift still elevates false alarm rates. DeepDefense~\cite{yuan2017}, Diro and Chilamkurti~\cite{diro2017}, and GraphSAGE-based approaches~\cite{hamilton2017} each demonstrate domain-tailored value but share a common limitation: evaluation on narrow single-dataset benchmarks that does not address the cross-capture-tool feature alignment.

\subsection{Multi-Dataset Evaluation and Feature Alignment}
\label{subsec:rel_multidataset}

Ring et al.~\cite{ring2019} found that single-dataset evaluation dominates the IDS literature, with naming inconsistencies and capture-tool variation being the primary obstacles for principled multi-dataset comparison. Engelen et al.~\cite{engelen2021} audited CICIDS-2017 and catalogued labelling errors and CICFlowMeter artefacts that inflates the reported metrics, reinforcing that high single-dataset F1 does not imply real-world generalisation.

To the best of our knowledge, no prior IDS work defines a formal, named canonical feature vocabulary with explicitly disclosed coverage statistics and genuine-equivalence-only mapping constraints applied simultaneously across five structurally distinct datasets. Existing multi-dataset approaches either restricts the evaluation to datasets sharing the same capture tool~\cite{ring2019}, apply PCA~\cite{jolliffe2002} to discard semantic interpretability, or employ ad-hoc name matching without any auditing semantic equivalence. To the best of our knowledge, \bridge{} represents the first formally specified and fully disclosed cross-dataset feature alignment and evaluation methodology for heterogeneous IoT network security datasets.

\subsection{Attention Mechanisms for Network Security}
\label{subsec:rel_attention}

Attention mechanisms have been applied in IDS to re-weight temporal steps in recurrent models~\cite{akuthota2025}. CB-GAF extends cross-attention in two key respects: it operates across three branches simultaneously (each querying the other two branches simultaneously), and a learnable sigmoid vector gate per branch enables feature-wise suppression of cross-branch information when it is unhelpful. It is a capability that is absent from vanilla cross-attention. This gating is particularly important in the heterogeneous setting, where branch informativeness varies with the canonical vocabulary coverage of the source dataset.

\subsection{Positioning of This Work}
\label{subsec:rel_positioning}

Table~\ref{tab:comparison} summarises the key dimensions on which TCH-Net is positioned relative to prior work. Table~\ref{tab:comparison} highlights that TCH-Net evaluated on \bridge{} is the only system combining a formally named multi-dataset benchmark, principled feature alignment, gated multi-branch fusion, and comprehensive evaluation including LODO generalization.

\begin{table}[t]
\centering
\caption{Qualitative comparison with representative prior work.
(\checkmark)~=~supported; (---)~=~not supported.}
\label{tab:comparison}
\resizebox{\columnwidth}{!}{%
\begin{tabular}{lccccc}
\toprule
\textbf{Method} &
\textbf{\shortstack{Multi-\\branch}} &
\textbf{\shortstack{Gated\\Fusion}} &
\textbf{\shortstack{Multi-\\dataset}} &
\textbf{\shortstack{IoT\\Focus}} &
\textbf{\shortstack{Full\\Ablation}} \\
\midrule
RF / XGB \cite{breiman2001,chen2016xgboost}
  & --- & --- & --- & --- & --- \\
BiLSTM \cite{imrana2021}
  & --- & --- & --- & \checkmark & --- \\
Transformer \cite{akuthota2025}
  & --- & --- & --- & \checkmark & --- \\
Kitsune \cite{mirsky2018}
  & \checkmark & --- & --- & \checkmark & --- \\
N-BaIoT \cite{meidan2018}
  & --- & --- & --- & \checkmark & --- \\
DeepDefense \cite{yuan2017}
  & --- & --- & --- & \checkmark & --- \\
IoT-DNN \cite{diro2017}
  & --- & --- & --- & \checkmark & --- \\
\textbf{TCH-Net (Ours)}
  & \checkmark & \checkmark & \checkmark & \checkmark & \checkmark \\
\bottomrule
\end{tabular}}
\end{table}

\section{BRIDGE: Datasets, Feature Alignment, and Preprocessing}
\label{sec:datasets}

A principled multi-dataset benchmark demands careful attention to three interrelated problems: the selection and characterisation of constituent datasets, the alignment of their heterogeneous feature spaces into a common representation, and the construction of a preprocessing pipeline that is transparent, reproducible, and free of data leakage. This section addresses each of these problems in turn.

\subsection{Dataset Selection Rationale}
\label{subsec:data_selection}

Five publicly available network security datasets are incorporated into \bridge{}, selected to cover the widest feasible range of capture modalities, network environments, device populations, and attack categories relevant to IoT botnet detection (Table~\ref{tab:datasets}). Crucially, each dataset was chosen because it fills a specific gap in the evaluation space that no other selected dataset covers.
This deliberate diversity is precisely what makes \bridge{} informative, that if all datasets shared the same capture tool and device environment, the benchmark would not stress feature alignment, and LODO results would not surface the cross-dataset domain shift that we have shown, which is a primary open challenge.

Datasets split into two tiers by coverage (Table~\ref{tab:datasets}): three \emph{primary} ($\geq$39\%) and two \emph{supplementary} ($\leq$22\%), the latter stress-testing generalisation across structurally distant feature spaces.

\begin{table}[t]
\centering
\caption{Overview of the five \bridge{} datasets.
$\star$ = Supplementary; low coverage reflects non-flow-level capture.}
\label{tab:datasets}
\resizebox{\columnwidth}{!}{%
\begin{tabular}{llcccc}
\toprule
\textbf{Dataset} & \textbf{Capture Tool} & \textbf{Year} &
\textbf{Coverage} & \textbf{Tier} \\
\midrule
CICIDS-2017    & CICFlowMeter & 2017 & 93\% & Primary \\
CIC-IoT-2023   & CICFlowMeter & 2023 & 87\% & Primary \\
Bot-IoT        & Argus        & 2019 & 39\% & Primary \\
Edge-IIoTset$^\star$ & Wireshark & 2022 & 22\% & Supplementary \\
N-BaIoT$^\star$      & Kitsune   & 2018 & 15\% & Supplementary \\
\bottomrule
\end{tabular}}
\end{table}

\subsection{Individual Dataset Characterisation}
\label{subsec:data_chars}

\subsubsection{CICIDS-2017}
\label{subsubsec:cicids2017}

CICIDS-2017~\cite{sharafaldin2018} achieves 93\% canonical vocabulary coverage and contributes approximately 28\% of post-balancing training records; 43 of 46 canonical slots receive genuine CICFlowMeter matches. It covers 14 attack types over a five-day testbed. Despite being well-documented labelling artefacts ~\cite{engelen2021} that inflate single-dataset metrics, we retain it as a calibration anchor: the multi-dataset evaluation prevents over-reliance on a single source, and its LODO result (Section~\ref{subsec:res_lodo}) explicitly quantifies non-transferable performance.

\subsubsection{CIC-IoT-2023}
\label{subsubsec:ciciot2023}

CIC-IoT-2023~\cite{neto2023} was built around 105 physical IoT devices tested under 18 MITRE ATT\&CK scenarios, producing traffic that reflects the constrained, bursty behaviour of embedded IoT firmware. CICFlowMeter capture gives 40/46 canonical matches (87\%). Its 2023 collection date makes it the most temporally proximate benchmark for current IoT threats in the suite.

\subsubsection{Bot-IoT}
\label{subsubsec:botiot}

Bot-IoT~\cite{koroniotis2018} was captured with Argus rather than CICFlowMeter, which is a session-level tool that exports byte counts, session duration, and TCP flags but not per-direction flow rates or subflow statistics, giving 39\% canonical coverage (18/46). Its attack scenarios are botnet-specific (DDoS, C\&C beaconing, exfiltration, reconnaissance), and the Argus/CICFlowMeter tool boundary is precisely the cross-capture-tool heterogeneity the vocabulary is designed to bridge. Its 38 post-balancing test samples preclude a reliable per-dataset metrics, which are excluded accordingly; Bot-IoT's value is structural, it is the only source imposing a 61\% zero-fill regime on the canonical vocabulary.

\subsubsection{Edge-IIoTset}
\label{subsubsec:edge}

Edge-IIoTset~\cite{ferrag2022} records packet-level traffic via Wireshark on Raspberry~Pi IIoT nodes running MQTT, Modbus, CoAP, DNP3, and AMQP; Wireshark operates below the flow-aggregation layer, so canonical coverage falls to 22\%~(10/46), filled only by inter-packet times, packet lengths, TCP flags, and header size. Its value lies in traffic character: IIoT protocols impose strict timing regularity that attacks disrupt in ways that differ sharply from IT-network intrusions, stress-testing generalisation to an environment structurally unlike the CICFlowMeter-dominated training corpus.

\subsubsection{N-BaIoT}
\label{subsubsec:nbaiot}

N-BaIoT~\cite{meidan2018} contains pre-computed Kitsune statistical fingerprints~\cite{mirsky2018}, 115-dimensional vectors with no direct CICFlowMeter correspondence, giving the lowest canonical coverage at 15\%~(7/46). Despite this, it still achieves the highest per-dataset F1: Mirai and BASHLITE infections produce stereotyped, high-volume traffic separable from benign behaviour even in just seven features. It also provides confirmed ground-truth labels from physical device compromises, making it a validity anchor for the detection task.

\subsection{Canonical Feature Vocabulary}
\label{subsec:vocab}

\subsubsection{Design Principles}
\label{subsubsec:vocab_design}

Three explicit constraints govern the vocabulary. \emph{Genuine equivalence only}: a feature maps to a canonical slot only if it measures the same network-theoretic quantity with the same computational definition, regardless of capture tool; superficially similar but semantically distinct quantities are not mapped. \emph{Explicit zero-filling}: absent features are set to zero for all records from the concerning dataset, making coverage gaps auditable. \emph{No dimensionality reduction}: PCA and similar projections are excluded as they destroys the  semantic interpretability.

\subsubsection{Vocabulary Structure}
\label{subsubsec:vocab_structure}

The 46 canonical features are organised into four semantically coherent groups (Table~\ref{tab:vocab_groups}). Groups~1 and~2 serve as primary inputs to the Temporal and Statistical branches respectively; Groups~3 and 4 are shared across branches.

\begin{table*}[t]
\centering
\caption{46-feature canonical vocabulary by semantic group.}
\label{tab:vocab_groups}
\begin{tabular}{clcc}
\toprule
\textbf{Group} & \textbf{Semantic Category} & \textbf{Indices} &
\textbf{Count} \\
\midrule
1 & Flow rates, durations, pkt/byte counts & 0--16  & 17 \\
2 & Packet size \& IAT statistics          & 17--37 & 21 \\
3 & TCP flag indicators                    & 38--43 &  6 \\
4 & Header length \& window size           & 44--45 &  2 \\
\midrule
  & \textbf{Total}                         &        & \textbf{46} \\
\bottomrule
\end{tabular}
\end{table*}

Group~1 captures temporal flow dynamics: duration, forward and backward packet/byte counts, per-direction rates, total flow rates, and subflow packet counts, encoding how a session evolves over time.
Group~2 captures statistical distributional structure: minimum, maximum, mean, and the standard deviation of packet lengths and inter-arrival times~(IATs) in both directions, particularly informative for distinguishing device classes. Groups~3 and~4 encode protocol-level signalling: individual TCP flag counts (SYN, ACK, FIN, RST, PSH, URG), forward header length, and initial window size.

\subsubsection{Per-Dataset Coverage}
\label{subsubsec:vocabcoverage}

\begin{figure*}[!t]
\centering
\includegraphics[width=\textwidth]{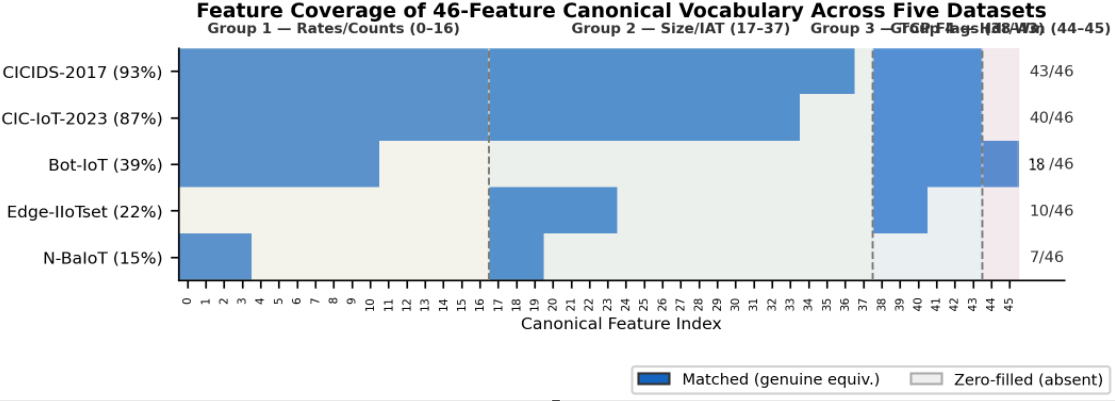}
\caption{Feature coverage of the 46-feature canonical vocabulary
  across five \bridge{} datasets. Blue cells indicate a genuinely matched feature; grey cells
  indicate explicit zero-fill (feature absent from that dataset). Column groups correspond to the four semantic categories in Table~\ref{tab:vocab_groups}. Coverage percentages are reported in
  Table~\ref{tab:vocab_coverage}.}
\label{fig:feature_coverage}
\end{figure*}

Table~\ref{tab:vocab_coverage} reports per-dataset matched feature counts and coverage
percentages. A feature is counted as matched only if an authentic semantic equivalent exists and
is verified by the alias mapping procedure. Figure~\ref{fig:feature_coverage} provides a visual representation of matched and zero-filled features across all five datasets.

\begin{table*}[t]
\centering
\caption{Canonical vocabulary coverage per dataset.
$\star$ = Non-flow-level capture.}
\label{tab:vocab_coverage}
\begin{tabular}{lccl}
\toprule
\textbf{Dataset} & \textbf{Matched / 46} & \textbf{Coverage} &
\textbf{Active Groups} \\
\midrule
CICIDS-2017    & 43 & 93\% & All four \\
CIC-IoT-2023   & 40 & 87\% & All four \\
Bot-IoT        & 18 & 39\% & Groups 1, 3, 4 \\
Edge-IIoTset$^\star$ & 10 & 22\% & Groups 2--4 (partial) \\
N-BaIoT$^\star$      &  7 & 15\% & Groups 1--2 (partial) \\
\bottomrule
\end{tabular}
\end{table*}

\subsubsection{Alias Mapping Procedure}
\label{subsubsec:alias}

Alignment uses a per-dataset alias map with three priority stages: exact case-insensitive match, alias exact match, and alias substring match ($\geq$5 characters). When multiple columns match, the highest-priority match is selected and flagged for auditing. Full mapping tables have been provided as a supplementary material.

\subsection{Preprocessing Pipeline}
\label{subsec:preprocessing}

\subsubsection{Class Balancing}
Records are separated by label and are subsampled to a 1:1 benign-to-attack ratio. This strict balance ensures no class dominates the training loss. The 1:1 ratio was selected after pilot experiments with 3:1 and 1:3 ratios revealed class collapse in datasets with low initial attack proportions (e.g., CICIDS-2017 at 14.5\%), where even a 1:3 oversampling produced window
attack incidence below 10\%. A minimum of 5,000 samples per class are preserved to prevent degenerate splits.

\subsubsection{Semantic Vector Construction}
Each record is mapped to the 46-dimensional canonical vector through alias
mapping. All values are parsed as 32-bit floats. Non-numeric, NaN, and infinite values are replaced with zero, producing a matrix $\mathbf{X}^{(d)} \in \mathbb{R}^{N_d \times 46}$ per dataset~$d$.

\subsubsection{Normalisation}
The five per-dataset matrices are concatenated to form $\mathbf{X}_{\mathrm{train}} \in \mathbb{R}^{N \times 46}$. A \texttt{RobustScaler}, centring by median, scaling by the 5th to the 95th percentile
interquartile range, and is fitted \emph{exclusively} on $\mathbf{X}_{\mathrm{train}}$ and applied to $\mathbf{X}_{\mathrm{test}}$ without refitting. Scaled values are clipped to $[-10, 10]$. A single shared scaler was deliberately used: per-dataset scaling would normalise away inter-dataset distributional differences that carry useful discriminative information and would constitute a form of data leakage in the LODO protocol.

\subsubsection{Sequence Construction}
A sliding window of length $W = 32$ and stride $S = 4$ was applied to each
dataset's records after being sorted by flow arrival time, producing sequence tensors
of shape $(N_{\mathrm{seq}}, 32, 46)$.
Window labels are assigned by majority vote over constituent record labels.
Training sequences are capped at 800{,}000 and test sequences at 200{,}000 for computational tractability on standard commodity hardware.

\subsubsection{Context Vector Construction}
Each sequence window receives an integer context vector $\mathbf{c} = (c_{\mathrm{ds}}, c_{\mathrm{dev}})$, where $c_{\mathrm{ds}} \in \{0,1,2,3,4\}$ identifies the source dataset and $c_{\mathrm{dev}} \in \{0,\ldots,5\}$ identifies the inferred device category. These identifiers serve as inputs to the Contextual branch.

\subsubsection{Train/Test Split and Leakage Verification}
The combined sequence dataset is split in a ratio of 80:20 by stratified random sampling. Splitting is performed \emph{after} sequence construction which prevents label leakage from windows spanning the split boundary.
Three data leakage verification checks were applied and all passed:
(i)~scaler fitted before any test-set access;
(ii)~hash-based overlap detection confirming zero identical feature vectors between train and test partitions;
(iii)~benign/attack ratio consistent between train~(0.758) and test~(0.750). Table~\ref{tab:records} reports post-balancing record counts.

\begin{table}[t]
\centering
\caption{Post-balancing record counts per dataset.}
\label{tab:records}
\begin{tabular}{lrrrr}
\toprule
\textbf{Dataset} & \textbf{Benign} & \textbf{Attack} & \textbf{Total} &
\textbf{Atk\%} \\
\midrule
CICIDS-2017   & 19,321 & 14,350 & 33,671 & 42.6\% \\
CIC-IoT-2023  &  3,964 &  3,001 &  6,965 & 43.1\% \\
Bot-IoT       &     22 &     16 &     38 & 42.1\% \\
Edge-IIoTset  & 30,951 & 23,435 & 54,386 & 43.1\% \\
N-BaIoT       & 13,557 & 10,055 & 23,612 & 42.6\% \\
\midrule
\textbf{Combined} & \textbf{67,815} & \textbf{50,857} &
\textbf{118,672} & \textbf{42.9\%} \\
\bottomrule
\end{tabular}
\end{table}

\section{Proposed Architecture: TCH-Net}
\label{sec:architecture}

\begin{figure*}[t]
    \centering
    \includegraphics[width=\textwidth, height=0.42\textheight]{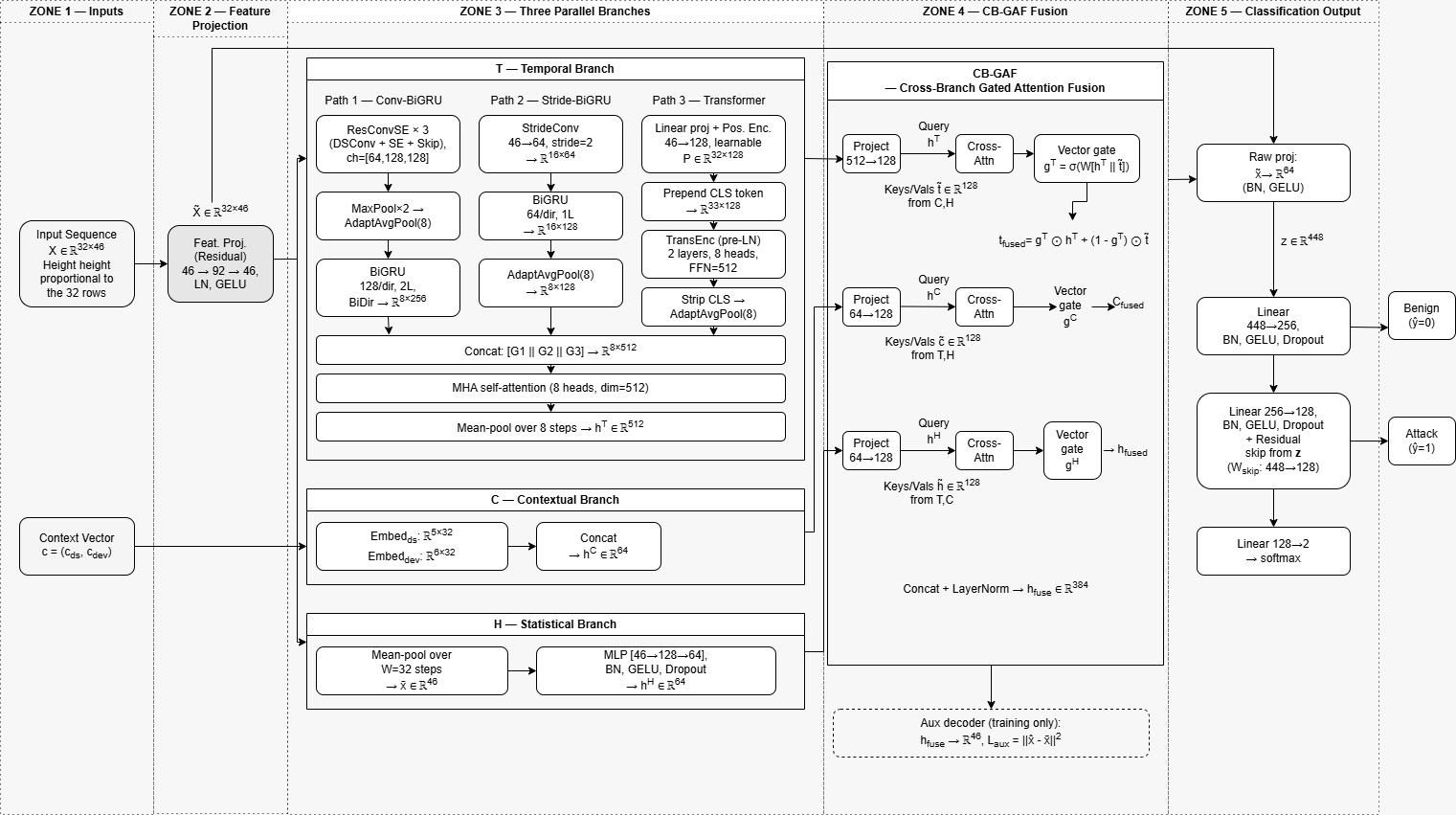}
    \caption{Full TCH-Net architecture overview across five zones: inputs, shared feature projection, three parallel branches (T, C, H), CB-GAF fusion, and classification output.}
    \label{fig:fig4}
\end{figure*}

TCH-Net is a multi-branch neural architecture that processes a sequences of canonical network flow feature vectors to produce a binary intrusion detection decisions. Three specialised parallel branches, the Temporal (T), Contextual (C), and Statistical (H) branches are preceded by a shared residual feature projection module and integrated by the Cross-Branch Gated Attention Fusion (CB-GAF) mechanism. A residual classification head and an auxiliary reconstruction decoder complete the model. The complete architecture is illustrated in Figure~\ref{fig:fig4}.
 
\subsection{Problem Formulation}
 
Let $\mathbf{X} = [\mathbf{x}_1, \ldots, \mathbf{x}_W] \in \mathbb{R}^{W \times F}$ denote a sequence of $W = 32$ consecutive canonical flow feature vectors, each of dimension $F = 46$. Let $\mathbf{c} = (c_\text{ds}, c_\text{dev}) \in \mathbb{Z}^2$ denote the context vector, where $c_\text{ds} \in \{0,1,2,3,4\}$ identifies the source dataset and $c_\text{dev} \in \{0,\ldots,5\}$ identifies the inferred device category. The task is to learn $f_\theta : (\mathbf{X}, \mathbf{c}) \mapsto \hat{y} \in \{0, 1\}$, where $\hat{y} = 0$ (benign) and $\hat{y} = 1$ (attack).

\subsection{Shared Input Feature Projection}
 
Before branching, the raw canonical input $\mathbf{X}$ is passed through a shared residual feature projection module that learns the non-linear interactions among the 46 canonical features. Many discriminative signals in the network flow data are either ratios or products of raw statistics. For instance, bytes-per-packet or forward-to-backward rate ratios, that are not explicitly present in the canonical vocabulary. The feature projection module discovers such cross-feature relationships by applying a two-layered feed-forward network with a residual connection:
 
\begin{equation}
    \tilde{\mathbf{X}} = \mathbf{X} + f_\text{proj}(\mathbf{X}), \quad f_\text{proj}(\mathbf{X}) = \mathbf{W}_2 \cdot \text{GELU}(\text{LN}(\mathbf{W}_1 \mathbf{X}^\top))^\top
\end{equation}
 
Specifically, $f_\text{proj}$ consists of: Linear($46 \to 92$) $\to$ LayerNorm(92) $\to$ GELU $\to$ Dropout($\delta/2$) $\to$ Linear($92 \to 46$) $\to$ LayerNorm(46), applied independently at each time step. The residual connection $\tilde{\mathbf{X}} = \mathbf{X} + f_\text{proj}(\mathbf{X})$ preserves the original feature magnitudes while augmenting them with learned interaction terms. All three branches receive $\tilde{\mathbf{X}}$ as an input.

\subsection{Temporal Branch (T): Three-Path Multi-Scale Temporal Encoding (MSTE)}

The T-branch captures the sequential dependencies across all three distinct temporal and temporal scales simultaneously. The central architectural motivation is that different botnet attack categories manifest at qualitatively distinct temporal scales: DDoS flooding produces discriminative burst-level signatures detectable within a few consecutive flows; C\&C beaconing produces medium-scale periodic patterns spanning tens of flows; and coordinated scan-then-exploit sequences produce a global ordering constraints across the entire 32-step window. A single-resolution encoder must trade sensitivity at one scale against the others. The T-branch resolves this by routing the input through three specialised parallel paths whose outputs are subsequently unified via multi-head self-attention over a shared temporal grid of 8 steps.
 
\begin{figure*}[t]
    \centering
    \includegraphics[width=\textwidth, height=0.38\textheight]{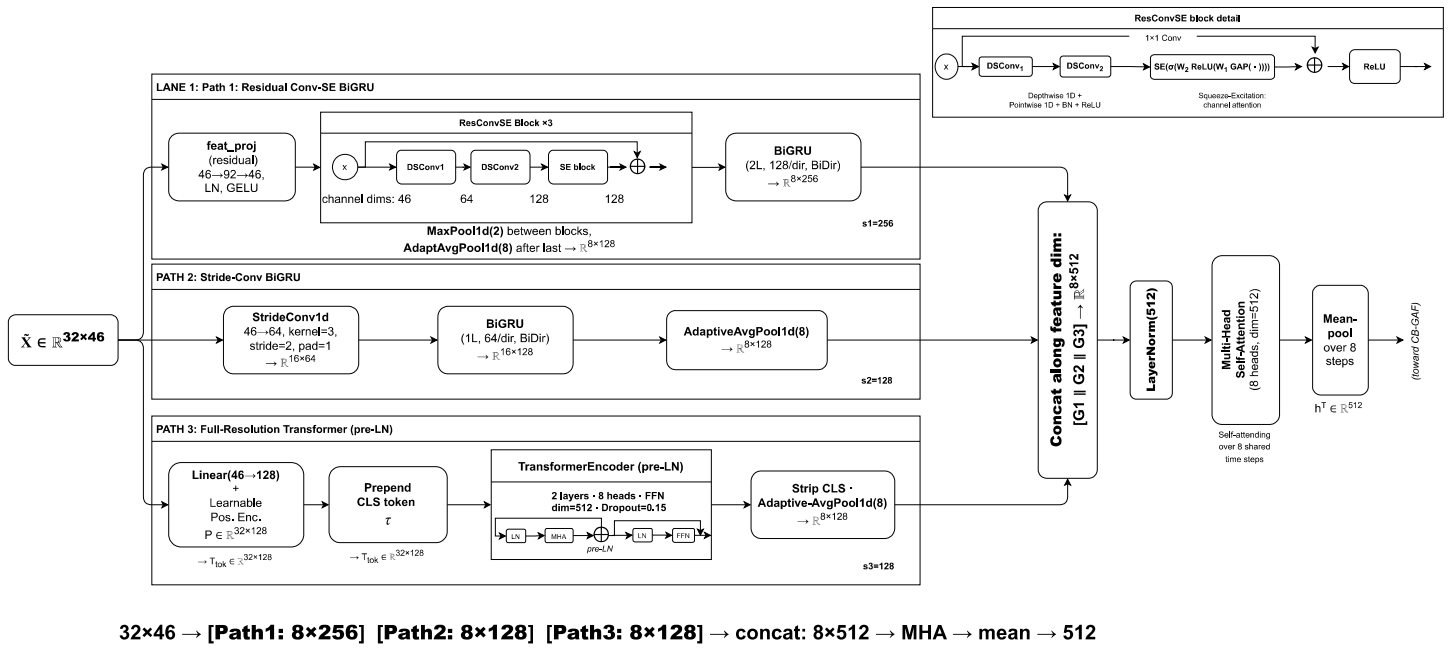}
    \caption{T-branch three-path architecture detail. Path 1: Residual Conv-SE BiGRU (local and medium-range patterns). Path 2: Stride-Conv BiGRU (coarse-scale patterns). Path 3: Full-resolution pre-LayerNorm Transformer (global temporal context). All three paths merge onto a shared 8-step temporal grid before multi-head self-attention and mean-pooling to produce $\mathbf{h}^T \in \mathbb{R}^{512}$.}
    \label{fig:fig1}
\end{figure*}
 
\subsubsection{Path 1: Residual Depthwise-Separable Convolutional BiGRU (Local and Medium-Range Patterns)}
 
Path 1 applies a three-stage convolutional frontend which is implemented as a stack of Residual Depthwise-Separable Convolutional blocks with Squeeze-Excitation recalibration (ResConvSE) and then followed by a two-layer bidirectional GRU.
 
\paragraph{Depthwise-Separable Convolution.} Each convolutional layer applies a depthwise convolution (one filter per input channel) followed by a pointwise convolution ($1\times1$ cross-channel mixing), reducing parameter count relative to standard convolution while preserving expressive capacity:
 
\begin{equation}
    \text{DSConv}(\mathbf{u}) = \text{ReLU}(\text{BN}(\mathbf{W}_\text{pw} \star (\mathbf{W}_\text{dw} \star \mathbf{u})))
\end{equation}
 
where $\star$ denotes convolution, $\mathbf{W}_\text{dw} \in \mathbb{R}^{C_\text{in} \times 1 \times k}$ is the depthwise filter (kernel width $k=3$), and $\mathbf{W}_\text{pw} \in \mathbb{R}^{C_\text{out} \times C_\text{in} \times 1}$ is the pointwise filter.
 
\paragraph{Squeeze-Excitation Recalibration.} Each ResConvSE block applies channel-wise attention after the two DSConv layers to reweight feature maps by their global importance:
 
\begin{equation}
    \text{SE}(\mathbf{u}) = \mathbf{u} \odot \sigma\!\left(\mathbf{W}_2 \cdot \text{ReLU}(\mathbf{W}_1 \cdot \text{GAP}(\mathbf{u}))\right)
\end{equation}
 
where GAP denotes global average pooling over the time dimension, $\mathbf{W}_1 \in \mathbb{R}^{\lfloor C/r \rfloor \times C}$ and $\mathbf{W}_2 \in \mathbb{R}^{C \times \lfloor C/r \rfloor}$ with reduction ratio $r = 8$, and $\sigma$ is the sigmoid function.
 
\paragraph{ResConvSE Block.} The full residual block composes the two DSConv layers, SE recalibration, and a skip connection:
 
\begin{equation}
\begin{split}
    \text{ResConvSE}(\mathbf{u}) = \text{ReLU}\!\bigr(\text{SE}(\text{DSConv}_2(\text{DSConv}_1(\mathbf{u}))) +  \\ \text{BN}(\mathbf{W}_\text{skip}\mathbf{u})\bigr)
\end{split}
\end{equation}
 
where $\mathbf{W}_\text{skip}$ is a $1\times1$ projection (with batch normalisation) when input and output channel counts differ, and the identity otherwise.
 
\paragraph{Three-Stage Convolutional Frontend.} Three ResConvSE blocks are stacked with intermediate MaxPool1d(2) operations to progressively compress the temporal dimension:
 
\begin{align}
    \mathbf{U}_1 &= \text{MaxPool}(\text{ResConvSE}_{46 \to 64}(\tilde{\mathbf{X}}^\top)) \in \mathbb{R}^{64 \times 16}\\
    \mathbf{U}_2 &= \text{MaxPool}(\text{ResConvSE}_{64 \to 128}(\mathbf{U}_1)) \in \mathbb{R}^{128 \times 8}\\
    \mathbf{U}_3 &= \text{AdaptiveAvgPool}(\text{ResConvSE}_{128 \to 128}(\mathbf{U}_2), 8) \in \mathbb{R}^{128 \times 8}
\end{align}
 
where the input $\tilde{\mathbf{X}}$ is transposed to channel-first format ($F \times W$) for convolutional processing.
 
\paragraph{Two-Layer BiGRU.} The compressed temporal representation $\mathbf{U}_3^\top \in \mathbb{R}^{8 \times 128}$ is processed by a two-layer bidirectional GRU with $d_\text{gru} = 128$ units per direction:
 
\begin{equation}
    \mathbf{G}_1 = \text{BiGRU}_1(\mathbf{U}_3^\top) \in \mathbb{R}^{8 \times 256}
\end{equation}

\subsubsection{Path 2: Stride-Downsampled Convolutional BiGRU (Coarse-Scale Patterns)}
 
Path 2 applies a single strided convolution to produce a coarser temporal representation, then encodes it with a single-layer BiGRU. The stride-2 convolution performs both feature projection and temporal downsampling in one step, halving the sequence length from 32 to 16:
 
\begin{equation}
    \mathbf{V} = \text{ReLU}(\text{BN}(\mathbf{W}_\text{down} \star_2 \tilde{\mathbf{X}}^\top)) \in \mathbb{R}^{64 \times 16}
\end{equation}
 
where $\mathbf{W}_\text{down} \in \mathbb{R}^{64 \times 46 \times 3}$ is a strided convolution (kernel width 3, stride 2, padding 1). A single-layer BiGRU with $d_\text{gru}/2 = 64$ units per direction encodes the downsampled sequence:
 
\begin{equation}
    \mathbf{G}_2 = \text{BiGRU}_2(\mathbf{V}^\top) \in \mathbb{R}^{16 \times 128}
\end{equation}
 
To align Path 2 to the shared 8-step temporal grid established by Path 1, adaptive average pooling is applied over the time dimension:
 
\begin{equation}
    \mathbf{G}_2^{(8)} = \text{AdaptiveAvgPool}(\mathbf{G}_2^\top, 8)^\top \in \mathbb{R}^{8 \times 128}
\end{equation}

\subsubsection{Path 3: Full-Resolution Pre-LayerNorm Transformer (Global Temporal Context)}
 
Path 3 processes all 32 steps of $\tilde{\mathbf{X}}$ through a two-layer Transformer encoder, giving the T-branch the same global temporal receptive field as the Transformer-IDS baseline while complementing it with the local and coarse-scale representations from Paths 1 and 2. A linear projection and learnable positional encoding map the canonical features to a Transformer embedding dimension $d_T = 128$:
 
\begin{equation}
    \mathbf{T}_\text{tok} = \tilde{\mathbf{X}} \mathbf{W}_\text{proj}^\top + \mathbf{P} \in \mathbb{R}^{32 \times 128}
\end{equation}
 
where $\mathbf{W}_\text{proj} \in \mathbb{R}^{128 \times 46}$ and $\mathbf{P} \in \mathbb{R}^{32 \times 128}$ is a learnable positional encoding matrix. A classification CLS token $\boldsymbol{\tau} \in \mathbb{R}^{1 \times 128}$ is prepended:
 
\begin{equation}
    \mathbf{T}_\text{in} = [\boldsymbol{\tau} \,\|\, \mathbf{T}_\text{tok}] \in \mathbb{R}^{33 \times 128}
\end{equation}
 
A two-layer TransformerEncoder with pre-LayerNorm (norm\_first), 8 attention heads, feed-forward dimension 512, and dropout $\delta$ processes $\mathbf{T}_\text{in}$:
 
\begin{equation}
    \mathbf{T}_\text{out} = \text{TransEnc}(\mathbf{T}_\text{in}) \in \mathbb{R}^{33 \times 128}
\end{equation}
 
Pre-LayerNorm normalises inputs to each sub-layer before the sub-layer computation, which empirically accelerates convergence and reduces gradient variance compared to the post-LayerNorm formulation used in the Transformer-IDS baseline. The CLS token output is discarded and the remaining 32 token representations are aligned to the shared 8-step grid:
 
\begin{equation}
    \mathbf{G}_3^{(8)} = \text{AdaptiveAvgPool}(\mathbf{T}_\text{out}[1:,:]^\top, 8)^\top \in \mathbb{R}^{8 \times 128}
\end{equation}

\subsubsection{Multi-Path Merge and Self-Attention Refinement}
 
The three path outputs, all sharing the 8-step temporal grid are concatenated along the feature dimension to form the joint multi-scale representation:
 
\begin{equation}
    \mathbf{G}_\text{cat} = [\mathbf{G}_1 \,\|\, \mathbf{G}_2^{(8)} \,\|\, \mathbf{G}_3^{(8)}] \in \mathbb{R}^{8 \times d_T^*}
\end{equation}
 
where $d_T^* = s_1 + s_2 + s_3 = 256 + 128 + 128 = 512$.
 
Multi-head self-attention with $n_\text{heads} = 8$ is applied to $\mathbf{G}_\text{cat}$ after layer normalisation, enabling the three paths to attend to and reweight each other's temporal representations at each of the 8 shared time steps:
 
\begin{equation}
    \mathbf{A}, \_ = \text{MHA}(\text{LN}(\mathbf{G}_\text{cat}), \text{LN}(\mathbf{G}_\text{cat}), \text{LN}(\mathbf{G}_\text{cat})) \in \mathbb{R}^{8 \times 512}
\end{equation}
 
Mean-pooling over the 8 time steps yields the final T-branch representation:
 
\begin{equation}
    \mathbf{h}^T = \frac{1}{8}\sum_{t=1}^{8} \mathbf{A}_t \in \mathbb{R}^{512}
\end{equation}

\subsection{Statistical Branch (H): Aggregate Flow MLP}
 
The H-branch encodes the aggregate distributional profile of each input window via mean-pooling over the time dimension, collapsing temporal structure to expose the window-level statistical character:
 
\begin{equation}
    \bar{\mathbf{x}} = \frac{1}{W} \sum_{t=1}^{W} \tilde{\mathbf{x}}_t \in \mathbb{R}^{46}
\end{equation}
 
A two-layer MLP with GELU activations, batch normalisation, and dropout processes $\bar{\mathbf{x}}$:
 
\begin{equation}
  \mathbf{h}^H = \text{Dropout}(\text{GELU}(\text{BN}(\mathbf{W}_{H2} \cdot \text{Dropout}(\text{GELU}(\text{BN}(\mathbf{W}_{H1} \bar{\mathbf{x}})))))) \in \mathbb{R}^{64}
\end{equation}
 
where $\mathbf{W}_{H1} \in \mathbb{R}^{128 \times 46}$ and $\mathbf{W}_{H2} \in \mathbb{R}^{64 \times 128}$. The H-branch captures information that is invariant to temporal ordering exactly what the T-branch is least suited to encode. The mean-pooled representation is particularly informative for distinguishing device classes through Group 2 packet size and IAT statistics, and for detecting high-volume botnet floods that produce sustained distributional shifts regardless of their temporal pattern.

\subsection{Contextual Branch (C): Provenance-Conditioned Domain Embedding}
 
The Contextual branch provides CB-GAF with explicit structural context about the source of each input window, specifically, which dataset it originated from and what device category it represents. This branch does not independently classify network flows; its value emerges exclusively within the CB-GAF fusion module.
 
Dataset and device category identifiers are mapped to dense embeddings of dimension $d_e = 32$:
 
\begin{equation}
    \mathbf{e}_\text{ds} = \mathbf{E}_\text{ds}[c_\text{ds}] \in \mathbb{R}^{32}, \quad \mathbf{e}_\text{dev} = \mathbf{E}_\text{dev}[c_\text{dev}] \in \mathbb{R}^{32}
\end{equation}
 
where $\mathbf{E}_\text{ds} \in \mathbb{R}^{5 \times 32}$ and $\mathbf{E}_\text{dev} \in \mathbb{R}^{6 \times 32}$ are learned embedding matrices. The two embeddings are concatenated directly to form the C-branch representation:
 
\begin{equation}
    \mathbf{h}^C = [\mathbf{e}_\text{ds} \,\|\, \mathbf{e}_\text{dev}] \in \mathbb{R}^{64}
\end{equation}
 
No MLP is applied; the raw concatenated embedding is passed directly to CB-GAF. The C-branch alone achieves near-random classification performance (F1 $\approx$ 0.60, AUC $\approx$ 0.50), confirming that dataset and device identifiers do not independently predict attack labels. Its role is to condition CB-GAF's vector gates on the canonical vocabulary coverage profile of the source dataset, enabling the fusion mechanism to calibrate cross-branch information mixing accordingly.

\subsection{Cross-Branch Gated Attention Fusion (CB-GAF)}
 
CB-GAF integrates the three branch representations $\mathbf{h}^T \in \mathbb{R}^{512}$, $\mathbf{h}^C \in \mathbb{R}^{64}$, and $\mathbf{h}^H \in \mathbb{R}^{64}$ through a mechanism that allows each branch to selectively incorporate information from the other two. The degree of cross-branch information flow is controlled by a learned vector gate per branch, enabling fine-grained, feature-wise modulation.
 
\begin{figure*}[t]
    \centering
    \includegraphics[width=\textwidth, height=0.38\textheight]{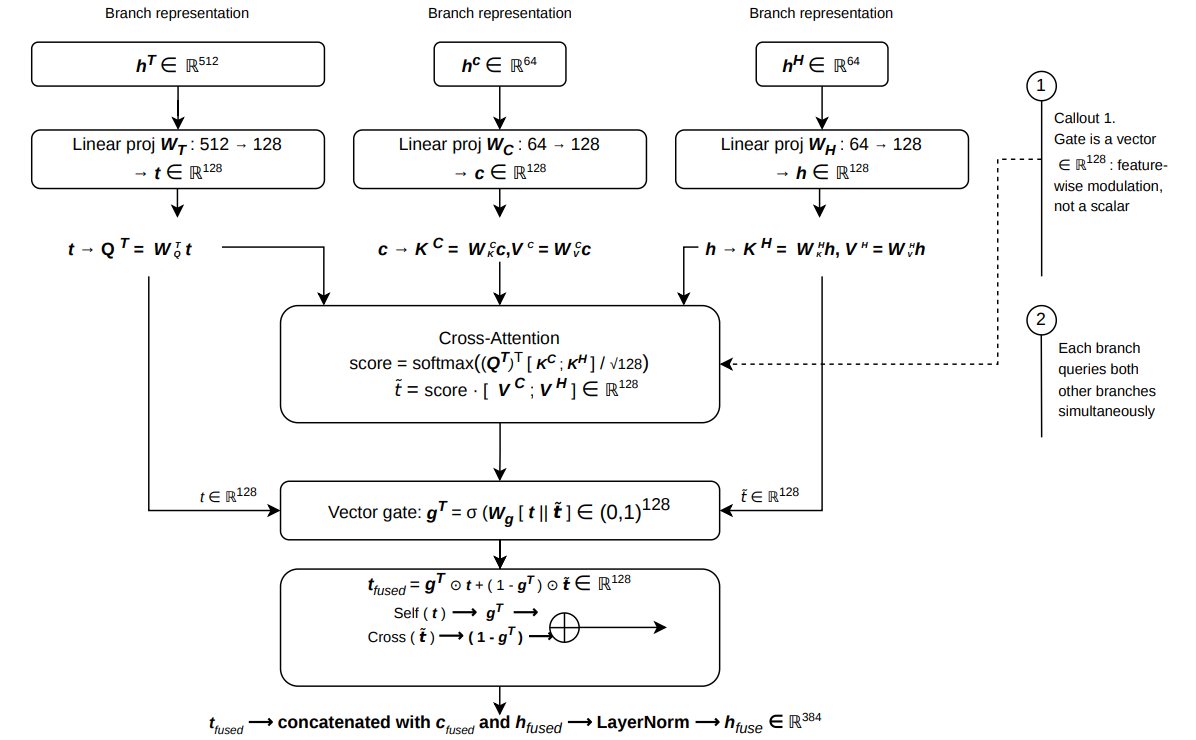}
    \caption{CB-GAF mechanism detail for branch $T$ as a representative example. Each branch projects to a common dimension $d_f = 128$, queries other two branches simultaneously via cross-attention, then applies a learned vector gate $\mathbf{g}^T \in (0,1)^{128}$ to produce the gated residual fusion $\mathbf{t}_\text{fused}$. Identical structure applied in parallel for branches C and H.}
    \label{fig:fig3}
\end{figure*}
 
\subsubsection{Branch Projection to Common Dimension}
 
Because the three branch representations have heterogeneous dimensionalities ($d_T^* = 512$, $d_C = d_H = 64$), each is first projected to a common fusion dimension $d_f = 128$ via learned linear maps:
 
\begin{equation}
    \mathbf{t} = \mathbf{W}_T \mathbf{h}^T \in \mathbb{R}^{128},\quad \mathbf{c} = \mathbf{W}_C \mathbf{h}^C \in \mathbb{R}^{128},\quad \mathbf{h} = \mathbf{W}_H \mathbf{h}^H \in \mathbb{R}^{128}
\end{equation}
 
where $\mathbf{W}_T \in \mathbb{R}^{128 \times 512}$, $\mathbf{W}_C \in \mathbb{R}^{128 \times 64}$, $\mathbf{W}_H \in \mathbb{R}^{128 \times 64}$.
 
\subsubsection{Cross-Branch Attention}
 
Each projected branch representation serves as a query attending simultaneously to the key-value pairs of the other two branches. For branch $T$ querying branches $C$ and $H$:
 
\begin{equation}
\begin{split}
    \mathbf{q}^T = \mathbf{W}_Q^T \mathbf{t}, \quad \mathbf{K}^T = [\mathbf{W}_K^C \mathbf{c} \,\|\, \mathbf{W}_K^H \mathbf{h}]^\top \in \mathbb{R}^{2 \times 128},   \\\\   \quad \mathbf{V}^T = [\mathbf{W}_V^C \mathbf{c} \,\|\, \mathbf{W}_V^H \mathbf{h}]^\top
\end{split}
\end{equation}
 
\begin{equation}
    \tilde{\mathbf{t}} = \text{softmax}\!\left(\frac{(\mathbf{q}^T)^\top \mathbf{K}^T}{\sqrt{d_f}}\right) \mathbf{V}^T \in \mathbb{R}^{128}
\end{equation}
 
The analogous operations for branches $C$ and $H$ are:
 
\begin{align}
    \tilde{\mathbf{c}} &= \text{Attn}(\mathbf{q}^C;\, \mathbf{K}^C = [\mathbf{W}_K^T \mathbf{t}, \mathbf{W}_K^H \mathbf{h}]) \in \mathbb{R}^{128}\\
    \tilde{\mathbf{h}} &= \text{Attn}(\mathbf{q}^H;\, \mathbf{K}^H = [\mathbf{W}_K^T \mathbf{t}, \mathbf{W}_K^C \mathbf{c}]) \in \mathbb{R}^{128}
\end{align}
 
where $\mathbf{W}_Q^i, \mathbf{W}_K^i, \mathbf{W}_V^i \in \mathbb{R}^{128 \times 128}$ for each branch $i \in \{T, C, H\}$.
 
\subsubsection{Learned Vector Gate and Residual Fusion}
 
A learnable sigmoid gate per branch controls the balance between the branch's own projected representation and the cross-attended signal. Crucially, the gate is a vector $\mathbf{g}^i \in (0,1)^{128}$ enabling feature-wise modulation of the fusion at each dimension independently. The gate is computed from the concatenation of the self-representation and the cross-attended output, allowing the gate to condition on both:
 
\begin{equation}
    \mathbf{g}^T = \sigma(\mathbf{W}_g^T [\mathbf{t} \,\|\, \tilde{\mathbf{t}}] + \mathbf{b}_g^T) \in (0,1)^{128}
\end{equation}
 
and analogously for $\mathbf{g}^C$ and $\mathbf{g}^H$, with $\mathbf{W}_g^i \in \mathbb{R}^{128 \times 256}$.
 
The gated residual fusion for each branch is:
 
\begin{equation}
    \mathbf{t}_\text{fused} = \mathbf{g}^T \odot \mathbf{t} + (\mathbf{1} - \mathbf{g}^T) \odot \tilde{\mathbf{t}} \in \mathbb{R}^{128}
\end{equation}
 
and analogously for $\mathbf{c}_\text{fused}$ and $\mathbf{h}_\text{fused}$. When $\mathbf{g}^i \to \mathbf{1}$, branch $i$ retains its own representation; when $\mathbf{g}^i \to \mathbf{0}$, it replaces its representation entirely with the cross-attended signal. This formulation is particularly critical in the heterogeneous multi-dataset setting: for inputs from low-coverage datasets (e.g., N-BaIoT at 15\% coverage), the H-branch is largely zero-padded; the gates on T and C can learn to down-weight H's contribution at the specific dimensions that are most affected, without hard-coding this decision and without sacrificing information from the remaining informative dimensions.
 
\subsubsection{Concatenation and Layer Normalisation}
 
The three gated branch outputs are concatenated and passed through a LayerNorm layer:
 
\begin{equation}
    \mathbf{h}_\text{fuse} = \text{LN}([\mathbf{t}_\text{fused} \,\|\, \mathbf{c}_\text{fused} \,\|\, \mathbf{h}_\text{fused}]) \in \mathbb{R}^{384}
\end{equation}

\subsection{Auxiliary Feature Reconstruction}
 
An auxiliary reconstruction objective prevents information collapse in CB-GAF during early training: a two-layer MLP decoder maps $\mathbf{h}_\text{fuse}$ back to the 46-dimensional canonical feature space:
 
\begin{equation}
    \hat{\mathbf{x}} = \mathbf{W}_{\text{dec},2} \cdot \text{GELU}(\mathbf{W}_{\text{dec},1} \mathbf{h}_\text{fuse}) \in \mathbb{R}^{46}
\end{equation}
 
\begin{equation}
    \mathcal{L}_\text{aux} = \frac{1}{F} \|\hat{\mathbf{x}} - \bar{\mathbf{x}}\|_2^2
\end{equation}
 
where $\mathbf{W}_{\text{dec},1} \in \mathbb{R}^{64 \times 384}$ and $\mathbf{W}_{\text{dec},2} \in \mathbb{R}^{46 \times 64}$. The decoder is discarded at inference time.

\subsection{Classification Head and Training Objective}
 
\subsubsection{Residual Classification Head}
 
A residual shortcut in the classification head improves the gradient flow. A raw feature projection from $\bar{\mathbf{x}}$ provides a direct low-level pathway:
 
\begin{equation}
    \mathbf{r} = \text{GELU}(\text{BN}(\mathbf{W}_\text{raw} \bar{\mathbf{x}})) \in \mathbb{R}^{64}
\end{equation}
 
where $\mathbf{W}_\text{raw} \in \mathbb{R}^{64 \times 46}$. The raw projection and the fused representation are concatenated to form the classifier input:
 
\begin{equation}
    \mathbf{z} = [\mathbf{h}_\text{fuse} \,\|\, \mathbf{r}] \in \mathbb{R}^{448}
\end{equation}
 
A two-layer MLP with a residual skip connection processes $\mathbf{z}$:
 
\begin{align}
    \mathbf{z}_1 &= \text{Dropout}(\text{GELU}(\text{BN}(\mathbf{W}_1 \mathbf{z}))) \in \mathbb{R}^{256}\\
    \mathbf{z}_2 &= \text{Dropout}(\text{GELU}(\text{BN}(\mathbf{W}_2 \mathbf{z}_1))) + \mathbf{W}_\text{skip} \mathbf{z} \in \mathbb{R}^{128}\\
    \hat{y} &= \text{softmax}(\mathbf{W}_\text{out} \mathbf{z}_2) \in \Delta^2
\end{align}
 
where $\mathbf{W}_1 \in \mathbb{R}^{256 \times 448}$, $\mathbf{W}_2 \in \mathbb{R}^{128 \times 256}$, $\mathbf{W}_\text{skip} \in \mathbb{R}^{128 \times 448}$ (the residual skip from input $\mathbf{z}$ directly to the second layer output), and $\mathbf{W}_\text{out} \in \mathbb{R}^{2 \times 128}$.
 
\begin{figure*}[t]
    \centering
    \includegraphics[width=\textwidth, height=0.30\textheight]{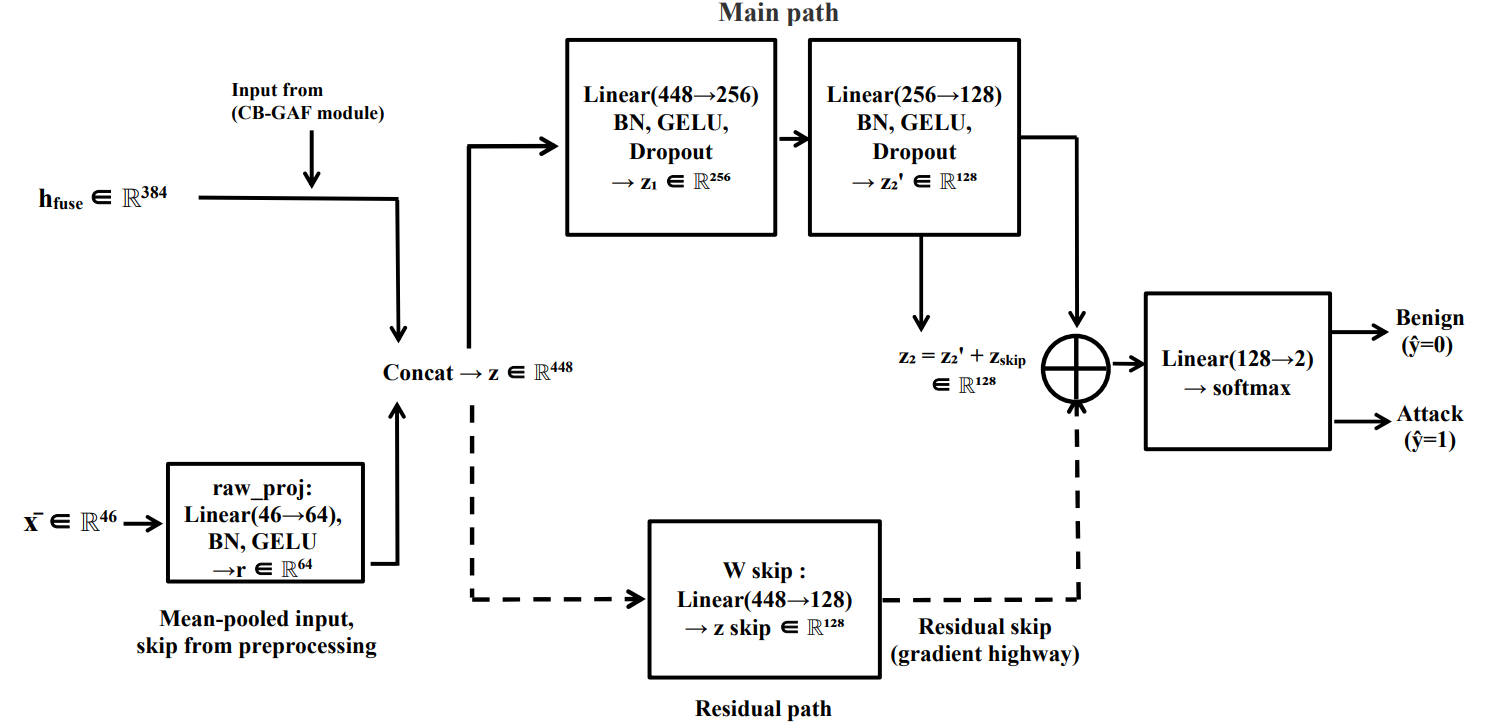}
    \caption{Classification head with residual skip connection detail. The mean-pooled input $\bar{\mathbf{x}} \in \mathbb{R}^{46}$ is projected to $\mathbf{r} \in \mathbb{R}^{64}$ and concatenated with $\mathbf{h}_\text{fuse}$ to form $\mathbf{z} \in \mathbb{R}^{448}$. A two-layer MLP with a residual skip $W_\text{skip}: 448 \to 128$ provides a direct gradient highway from the full input to the output layer, preventing gradient vanishing in the deep classification pathway.}
    \label{fig:fig2}
\end{figure*}
 
\subsubsection{Training Objective}
 
The total training loss combines focal classification loss with the auxiliary reconstruction term:
 
\begin{equation}
    \mathcal{L} = \mathcal{L}_\text{cls} + \lambda \mathcal{L}_\text{aux}, \quad \lambda = 0.05
\end{equation}
 
where $\mathcal{L}_\text{cls}$ is class-weighted focal loss with focusing parameter $\gamma = 2.0$ and label smoothing $\varepsilon = 0.05$:
 
\begin{equation}
    \mathcal{L}_\text{cls} = -\sum_{i} \alpha_i (1 - p_{t,i})^\gamma \log p_{t,i}
\end{equation}
 
Class weights $\alpha_i$ are set inversely proportional to class frequency in each training batch, normalised so that the mean weight equals 1. Label smoothing distributes $\varepsilon/2$ probability mass from each true class to the other, improving calibration.
 
\paragraph{Online Input Augmentation.} During training, zero-mean Gaussian noise with standard deviation $\sigma_\text{aug} = 0.010$ is added to each input sequence with probability $p_\text{aug} = 0.30$, after which values are clipped to $[-10, 10]$. This augmentation is applied exclusively during training and simulates sensor measurement noise, improving robustness to feature perturbation.

\subsection{Optimisation and Hyperparameters}
 
All models use AdamW with initial learning rate $\eta = 5 \times 10^{-4}$ and weight decay $\lambda_w = 5 \times 10^{-5}$. A cosine annealing schedule with 2-epoch linear warm-up is applied over a maximum of 30 epochs; early stopping is triggered after 5 epochs without validation F1 improvement. Table~\ref{tab:hyperparams} provides full hyperparameter details.
 
TCH-Net has 2.692M trainable parameters (2,691,696; verified programmatically from the released experimental pipeline). The approximate distribution across principal components is as follows: T-branch $\approx$1.98M (ResConvSE frontend $\approx$0.21M; BiGRU Path~1 $\approx$0.57M; Stride-BiGRU Path~2 $\approx$0.14M; Transformer Path~3 $\approx$0.43M; feat\_proj $\approx$0.035M; merge MHA $\approx$0.57M); C-branch $\approx$0.01M (embedding tables); H-branch $\approx$0.02M; CB-GAF $\approx$0.43M; classification head $\approx$0.18M; auxiliary decoder $\approx$0.02M. Per-component figures are approximate proportional estimates; the programmatically verified total is 2.692M, reported alongside the complete efficiency analysis in Section~\ref{sec:efficiency}.

TCH-Net's parameter count of 2.692M is approximately 4.4$\times$ that of the BiLSTM-IDS and Transformer-IDS baselines (0.609M and 0.618M respectively), reflecting the three-path T-branch design. This capacity overhead is contextualised by a measured single-sample inference latency of 6.43~ms on an NVIDIA Tesla T4 which is well within the range viable for edge inference accelerators, as examined quantitatively in Section~\ref{sec:efficiency}. To assess whether the performance gain is attributable to architectural design rather than raw capacity, we note that the T-branch ablation variant using only Path~1 (Conv-GRU alone, $\approx$2.1M parameters, approximately 3.4$\times$ the capacity of the BiLSTM-IDS baseline at 0.609M) achieves F1~$=$~0.7753, essentially matching the BiLSTM-IDS baseline (F1~$=$~0.7805). An architecture carrying 3.4$\times$ the parameter budget of the strongest recurrent baseline yet attaining near-identical detection performance when all architectural novelty is removed constitutes strong evidence that the gains from Paths~2 and~3 and CB-GAF fusion reflect genuine architectural contribution and not a capacity advantage. All baselines are re-evaluated on identical hardware and data pipeline to ensure consistent comparison conditions.

\begin{table*}[t]
\centering
\caption{Hyperparameter Settings}
\label{tab:hyperparams}
\renewcommand{\arraystretch}{1.25}
\begin{tabular}{lll}
\toprule
\textbf{Hyperparameter} & \textbf{Value} & \textbf{Rationale} \\
\midrule
Window $W$ & 32 & Covers short attack bursts \\
Stride $S$ & 4 & Coverage vs.\ compute \\
Embed dim $d_e$ & 32 & Compact context representation \\
Conv channels & [64, 128, 128] & Progressive feature expansion \\
SE reduction ratio $r$ & 8 & Channel recalibration efficiency \\
GRU hidden (Path 1) & 128/dir, 2 layers & Matched to baselines \\
GRU hidden (Path 2) & 64/dir, 1 layer & Coarse-scale, lower capacity \\
Transformer dim $d_T$ & 128 & Matched to Transformer-IDS baseline \\
Transformer layers & 2 & Pre-LN, 8 heads, FFN dim 512 \\
T-branch merged dim & 512 & $s_1 + s_2 + s_3 = 256+128+128$ \\
H-branch output dim & 64 & Compact aggregate representation \\
C-branch output dim & 64 & $2 \times d_e = 2 \times 32$ \\
CB-GAF proj dim $d_f$ & 128 & Expressivity / parameter balance \\
CB-GAF gate type & Vector ($\in \mathbb{R}^{128}$) & Feature-wise modulation \\
CB-GAF output dim & 384 & $3 \times d_f$ \\
Classifier input dim & 448 & $384 + 64$ (fused + raw\_proj) \\
Dropout $\delta$ & 0.15 & Applied to all paths \\
Noise aug.\ prob. & 0.30 & Training only \\
Noise aug.\ std. & 0.010 & Below scaler clip threshold \\
Epochs (max) & 30 & Full cosine schedule \\
Warm-up & 2 epochs & Smooth learning rate ramp \\
Early stopping & 5 epochs & Plateau detection on val F1 \\
Batch size & 512 & GPU utilisation \\
LR $\eta$ & $5 \times 10^{-4}$ & Stable with cosine annealing \\
Weight decay & $5 \times 10^{-5}$ & L2 regularisation \\
Focal $\gamma$ & 2.0 & Hard example focus \\
Label smooth $\varepsilon$ & 0.05 & Calibration floor \\
Aux weight $\lambda$ & 0.05 & Auxiliary $\ll$ primary loss \\
\bottomrule
\end{tabular}
\end{table*}

\section{Experimental Results}
\label{sec:results}

This section presents the complete experimental evaluation of TCH-Net across the seven components: (i)~setup and evaluation protocol; (ii)~baseline comparison; (iii)~branch ablation; (iv)~novelty component ablation; (v)~per-dataset performance breakdown; (vi)~temporal split evaluation; and (vii)~\bridge{} leave-one-dataset-out generalisation benchmark.

\subsection{Experimental Setup}
\label{subsec:res_setup}

\subsubsection{Hardware and Software}
All experiments are conducted on Kaggle Notebooks with NVIDIA Tesla T4 GPUs (16~GB VRAM), using PyTorch~2.x, scikit-learn~1.2, and XGBoost~1.7. This standardised cloud environment ensures that the results are reproducible on widely accessible commodity hardware.Inference latency is measured as the mean over $n=200$ single-sample forward passes following 20 GPU warm-up passes, timed using CUDA event synchronisation; throughput is reported as samples per second under batch-512 processing on the
same device. This protocol ensures that the efficiency figures reported in Section~\ref{sec:efficiency} are reproducible and free of cold-start bias.

\subsubsection{Evaluation Protocol}

\textsc{TCH-Net} results are reported as mean\,$\pm$\,std across five independent random seeds $\{42, 123, 456, 789, 2024\}$. Each seed performs fresh subsampling, splitting, and full training from scratch. Baseline models and branch ablation variants are evaluated over three seeds $\{42, 123, 456\}$. Novelty component ablation variants (\textsc{TCHNovAbl}) are evaluated over two seeds
$\{42, 123\}$. The LODO generalisation benchmark uses two seeds $\{42, 123\}$ per fold; the $5 \times 2 = 10$ full-training runs this entails preclude additional seeds under the available compute budget.

Statistical significance is assessed using the one-sided paired Wilcoxon signed-rank test~\cite{wilcoxon1992}, reported as $^*p < 0.05$, $^{**}p < 0.01$, $^{***}p < 0.001$.

\subsubsection{Metrics}
Four primary metrics are reported:
\textbf{F1 score}~(harmonic mean of precision and recall, robust to class
imbalance);
\textbf{ROC-AUC}~(threshold-independent discriminative ability);
\textbf{MCC}~(Matthews Correlation Coefficient; sensitive to all four confusion
matrix cells~\cite{chicco2020});
\textbf{PR-AUC}~(precision-recall curve area; particularly informative when the
attack class is primary).

\subsection{Baseline Models}
\label{subsec:res_baselines}

Twelve baselines are evaluated across five methodological families. All deep learning baselines use the same data pipeline, class balancing, normalisation, and sequence construction as TCH-Net.
Classical ML baselines operate on mean-pooled feature vectors.

\begin{enumerate}
\item \textbf{BiLSTM-IDS}~\cite{imrana2021}: Bidirectional LSTM, 128~units/direction, 2~layers, 32-step sequences.
\item \textbf{BiGRU-IDS}~\cite{cho2014}: Identical to BiLSTM-IDS using GRU cells.
\item \textbf{1D-CNN-IDS}: Three-layer 1D CNN, filters $[64, 128, 128]$, kernel width~3, global average pooling.
\item \textbf{Transformer-IDS}~\cite{akuthota2025}: 4-layer encoder, 8~heads, hidden dim~128, 32-step input.
\item \textbf{MLP-IDS}: Three-layer MLP on mean-pooled 46-dimensional vectors.
\item \textbf{CNN-LSTM}: Two 1D-CNN layers followed by bidirectional LSTM.
\item \textbf{Random Forest}~\cite{breiman2001}: 200~trees on mean-pooled vectors.
\item \textbf{XGBoost}~\cite{chen2016xgboost}: 200~estimators, max depth~6.
\item \textbf{Kitsune-AE}~\cite{mirsky2018}: Feature-group autoencoder ensemble, threshold~0.5.
\item \textbf{DeepDefense}~\cite{yuan2017}: Recurrent DDoS detector adapted to binary classification.
\item \textbf{GraphSAGE-Approx}~\cite{hamilton2017}: GraphSAGE neighbourhood aggregation on flow features.
\item \textbf{IoT-DNN}~\cite{diro2017}: Three-layer DNN with batch normalisation for IoT traffic.
\end{enumerate}

\subsection{Main Comparison Results}
\label{subsec:res_main}

Table~\ref{tab:main_results} reports the full comparison. TCH-Net achieves the highest score on all four primary metrics, outperforming every baseline with statistical significance ($p < 0.05$).

\begin{table*}[t]
\centering
\caption{Main comparison results. Mean~$\pm$~std across seeds.
Best per metric in \textbf{bold}.
$\Delta$F1: absolute F1 improvement of TCH-Net over each baseline.
Significance: $^{***}p < 0.001$, $^{**}p < 0.01$, $^{*}p < 0.05$
(one-sided paired Wilcoxon signed-rank test).}
\label{tab:main_results}
\resizebox{\textwidth}{!}{%
\begin{tabular}{lccccc}
\toprule
\textbf{Model} & \textbf{F1} & \textbf{ROC-AUC} & \textbf{MCC} &
\textbf{PR-AUC} & \textbf{$\Delta$F1 (sig.)} \\
\midrule
\textbf{TCH-Net (Ours)} &
  $\mathbf{0.8296 \pm 0.0028}$ &
  $\mathbf{0.9380 \pm 0.0025}$ &
  $\mathbf{0.6972 \pm 0.0056}$ &
  $\mathbf{0.8912 \pm 0.0031}$ & --- \\
\midrule
BiLSTM-IDS \cite{imrana2021}
  & $0.7805 \pm 0.0010$ & $0.8975 \pm 0.0001$ & $0.5972 \pm 0.0030$
  & $0.8441 \pm 0.0018$ & $+0.0491^{**}$ \\
BiGRU-IDS \cite{cho2014}
  & $0.7805 \pm 0.0011$ & $0.8962 \pm 0.0013$ & $0.5987 \pm 0.0034$
  & $0.8438 \pm 0.0022$ & $+0.0491^{**}$ \\
1D-CNN-IDS
  & $0.7932 \pm 0.0076$ & $0.9076 \pm 0.0062$ & $0.6213 \pm 0.0153$
  & $0.8601 \pm 0.0091$ & $+0.0364^{*}$ \\
Transformer-IDS \cite{akuthota2025}
  & $0.7958 \pm 0.0030$ & $0.9147 \pm 0.0012$ & $0.6255 \pm 0.0067$
  & $0.8699 \pm 0.0041$ & $+0.0338^{**}$ \\
MLP-IDS
  & $0.7039 \pm 0.0008$ & $0.8152 \pm 0.0005$ & $0.4348 \pm 0.0018$
  & $0.7311 \pm 0.0012$ & $+0.1257^{***}$ \\
CNN-LSTM
  & $0.7919 \pm 0.0137$ & $0.9056 \pm 0.0123$ & $0.6208 \pm 0.0261$
  & $0.8578 \pm 0.0167$ & $+0.0377^{*}$ \\
Random Forest \cite{breiman2001}
  & $0.4323 \pm 0.0082$ & $0.8005 \pm 0.0002$ & $0.3557 \pm 0.0043$
  & $0.6214 \pm 0.0053$ & $+0.3973^{***}$ \\
XGBoost \cite{chen2016xgboost}
  & $0.7265 \pm 0.0014$ & $0.8704 \pm 0.0002$ & $0.5542 \pm 0.0007$
  & $0.7989 \pm 0.0009$ & $+0.1031^{***}$ \\
Kitsune-AE \cite{mirsky2018}
  & $0.7045 \pm 0.0007$ & $0.8200 \pm 0.0001$ & $0.4362 \pm 0.0028$
  & $0.7348 \pm 0.0015$ & $+0.1251^{***}$ \\
DeepDefense \cite{yuan2017}
  & $0.7627 \pm 0.0011$ & $0.8776 \pm 0.0008$ & $0.5638 \pm 0.0039$
  & $0.8192 \pm 0.0027$ & $+0.0669^{***}$ \\
GraphSAGE-Approx \cite{hamilton2017}
  & $0.7097 \pm 0.0004$ & $0.8259 \pm 0.0003$ & $0.4465 \pm 0.0010$
  & $0.7403 \pm 0.0011$ & $+0.1199^{***}$ \\
IoT-DNN \cite{diro2017}
  & $0.7009 \pm 0.0002$ & $0.8146 \pm 0.0002$ & $0.4278 \pm 0.0017$
  & $0.7286 \pm 0.0009$ & $+0.1287^{***}$ \\
\midrule
\multicolumn{6}{l}{\textit{TCH-Net outperforms all 12 baselines on all metrics,
all statistically significant.}} \\
\bottomrule
\end{tabular}}
\end{table*}

\begin{figure}[!t]
\centering
\includegraphics[width=\columnwidth]{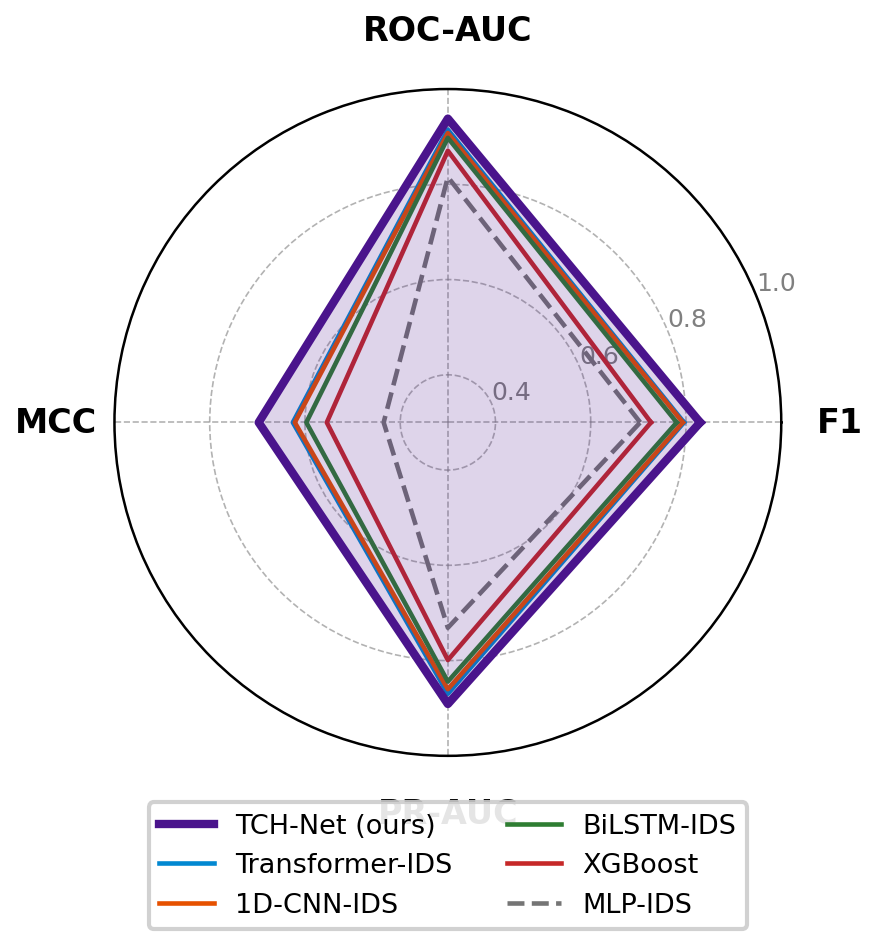}
\caption{Radar chart comparing TCH-Net against the five strongest
  baselines across all four evaluation metrics.
  TCH-Net (filled area) consistently extends beyond all
  baselines on every axis.}
\label{fig:radar}
\end{figure}

Figure~\ref{fig:radar} provides a radar chart comparing TCH-Net against the five strongest baselines across all four evaluation metrics. Among deep learning baselines, Transformer-IDS achieves the second-highest F1 $(0.7958)$ but the highest seed-to-seed variance, consistent with the known data-sensitivity of transformer models. 1D-CNN-IDS~$(0.7932)$ is competitive but limited to locally connected temporal regions. BiLSTM and BiGRU achieve near-identical F1~$(0.7805)$, confirming that the
performance gap is attributable to multi-branch fusion rather than recurrent cell selection.
Classical models~(Random Forest~$0.43$; XGBoost~$0.73$) show substantially lower F1 due to their inability to model temporal sequence structure.

\subsection{Computational Efficiency Analysis}
\label{sec:efficiency}

Table~\ref{tab:efficiency} reports the computational cost profile for TCH-Net and the five deep learning baselines, all benchmarked on NVIDIA Tesla T4 hardware under identical conditions. F1 values are the canonical figures from Table~\ref{tab:main_results} (Section~\ref{subsec:res_main}).

\begin{table*}[t]
\caption{Computational efficiency comparison. Latency: single-sample mean\,$\pm$\,std over $n=200$ runs following 20 GPU warm-up passes, CUDA event timing, NVIDIA Tesla T4. Throughput: batch-512 inference, same device. Mem: runtime GPU memory footprint. F1: canonical mean from Table~\ref{tab:main_results}. $\star$ = proposed model.}
\label{tab:efficiency}
\centering
\small
\setlength{\tabcolsep}{7pt}
\begin{tabular}{lrrrrr}
\toprule
\textbf{Model} &
\textbf{Params\,(M)} &
\textbf{Latency\,(ms)} &
\textbf{Throughput\,($\times$10$^3$\,sps)} &
\textbf{Mem\,(MB)} &
\textbf{F1} \\
\midrule
TCH-Net (Ours)$^\star$              & 2.692 & $6.43 \pm 0.18$ & 20.5  & 10.27 & \textbf{0.8296} \\
BiLSTM-IDS~\cite{imrana2021}        & 0.609 & $0.74 \pm 0.02$ & 34.2  &  2.32 & 0.7805 \\
BiGRU-IDS~\cite{cho2014}            & 0.465 & $0.58 \pm 0.03$ & 52.0  &  1.77 & 0.7805 \\
1D-CNN-IDS                          & 0.068 & $0.69 \pm 0.03$ & 406.9 &  0.26 & 0.7932 \\
Transformer-IDS~\cite{akuthota2025} & 0.618 & $1.22 \pm 0.03$ & 36.8  &  2.36 & 0.7958 \\
CNN-LSTM                            & 0.142 & $0.88 \pm 0.05$ & 273.5 &  0.54 & 0.7919 \\
\bottomrule
\end{tabular}
\end{table*}

In absolute terms, all evaluated models impose a negligible computational cost for IoT gateway deployment. TCH-Net's single-sample latency of 6.43\,ms corresponds to approximately 155 detection decisions per second in sequential mode and over 20,000 per second under batch-512 processing rates that comfortably accommodate continuous flow-level monitoring at an IoT gateway, where enterprise-class devices typically aggregate traffic at hundreds to a few thousand flows per second. The latency overhead relative to simpler baselines is therefore not a deployment barrier but a cost to be weighed against the F1 improvement it delivers.

TCH-Net achieves the highest F1 ($0.8296$) at the expense of higher latency and a 10.27\,MB memory footprint. Compared directly to BiLSTM-IDS, which is the strongest single-path recurrent baseline (F1\,=\,0.7805, latency 0.74\,ms, 2.32\,MB), and the trade-off is $+0.0491$ F1 for an approximately $8.7\times$ increase in per-sample latency and a $4.4\times$ larger footprint. In security-critical environments where detection quality is the primary objective, this trade-off favours TCH-Net. Deployments where latency is the binding constraint can use the lighter baselines in this suite as viable alternatives at a known F1 cost.

TCH-Net's 10.27\,MB footprint is readily accommodated on edge inference accelerators such as the NVIDIA Jetson family (Jetson Nano: 4\,GB LPDDR4; Jetson Orin NX: up to 16\,GB LPDDR5), but exceeds the on-chip SRAM of microcontroller-class endpoints (ARM Cortex-M, ESP32; typically below 1\,MB). Quantisation, structured pruning, and knowledge distillation~\cite{hinton2014distilling} represent well-established compression pathways for constrained targets, as discussed in Section~\ref{sec:edge}.

\subsection{Branch Ablation}
\label{subsec:res_branch}

\begin{figure}[!t]
\centering
\includegraphics[width=\columnwidth]{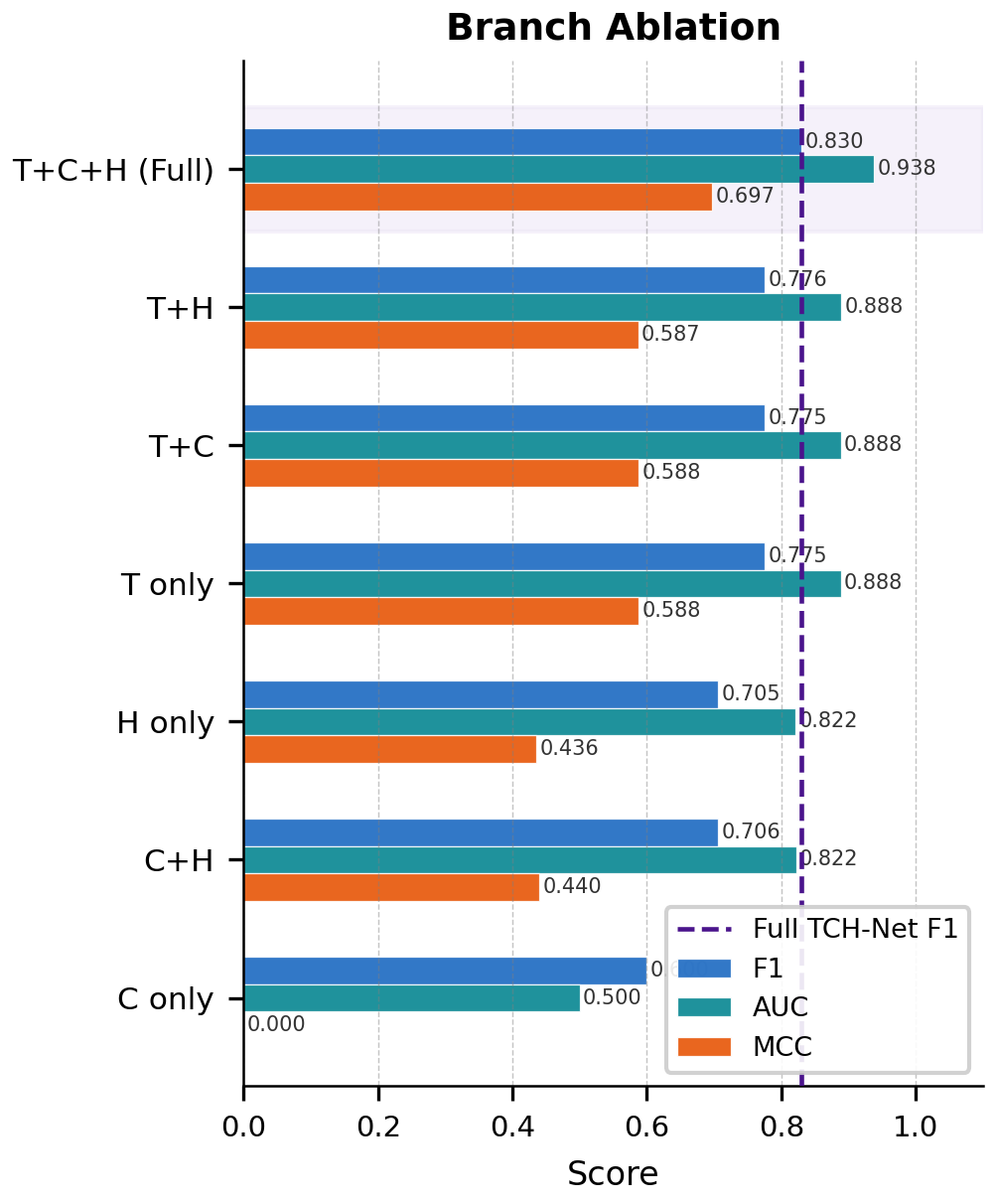}
\caption{Branch ablation F1, AUC, and MCC scores for all seven branch subsets. The dashed vertical line marks the full TCH-Net F1 ($= 0.8296$). All-branch fusion with \cbgaf{} provides the highest performance on every metric.}
\label{fig:branch_ablation_figure}
\end{figure}

All seven non-empty branch subsets are evaluated with CB-GAF fusion replaced by simple concatenation, reported over two seeds \{42, 123\}. Results are presented in Table~\ref{tab:branch_ablation_table} and Figure~\ref{fig:branch_ablation_figure}.

\begin{table}[t]
\centering
\caption{Branch ablation results (2 seeds each, seeds \{42, 123\}). Full model in bold.}
\label{tab:branch_ablation_table}
\begin{tabular}{lcccc}
\toprule
\textbf{Variant} & \textbf{F1~$\pm$~std} & \textbf{AUC} &
\textbf{MCC} & \textbf{$\Delta$F1} \\
\midrule
\textbf{TCH (Full)} & $\mathbf{0.8296 \pm 0.0028}$ & \textbf{0.9380} &
  \textbf{0.6972} & --- \\
T+C  & $0.7752 \pm 0.0012$ & 0.8883 & 0.5880 & $-0.0544$ \\
T+H  & $0.7756 \pm 0.0014$ & 0.8885 & 0.5870 & $-0.0540$ \\
C+H  & $0.7061 \pm 0.0003$ & 0.8218 & 0.4402 & $-0.1235$ \\
T only & $0.7753 \pm 0.0013$ & 0.8882 & 0.5878 & $-0.0543$ \\
C only & $0.6000 \pm 0.0000$ & 0.4999 & 0.0000 & $-0.2296$ \\
H only & $0.7054 \pm 0.0003$ & 0.8215 & 0.4359 & $-0.1242$ \\
\bottomrule
\end{tabular}
\end{table}

The full \textsc{TCH-Net} outperforms all branch-subset variants by a substantial margin ($+0.054$ F1 over T+H, the best two-branch proxy combination), with the gap reflecting both branch removal and the contribution of Path~3 and \texttt{feat\_proj} absent from proxy variants. The C-branch alone achieves near-random performance~(AUC~$\approx 0.50$, MCC~$= 0.000$), confirming that dataset identifiers do not predict attack labels
independently, consistent with the C-branch's role as a provenance conditioning signal for CB-GAF rather than an independent classifier. T alone outperforms H alone~$(0.7753$ vs.\ $0.7054)$, consistent with temporal structure being more discriminative than aggregate statistics for sequential attack detection; yet T+H exceeds T alone, confirming genuine complementarity.

\subsection{Novelty Component Ablation}
\label{subsec:res_novelty}

Table~\ref{tab:novelty_ablation} presents four variants selectively removing each
novel component.

\begin{table*}[t]
\centering
\begin{threeparttable}
\caption{Novelty component ablation (2~seeds, $\{42, 123\}$). Full model in bold.}
\label{tab:novelty_ablation}
\begin{tabular}{lcccc}
\toprule
\textbf{Variant} & \textbf{F1~$\pm$~std} & \textbf{AUC} &
\textbf{MCC} & \textbf{$\Delta$F1} \\
\midrule
\textbf{Full TCH-Net} & $\mathbf{0.8296 \pm 0.0028}$ & \textbf{0.9380} &
  \textbf{0.6972} & --- \\
w/o CB-GAF    & $0.7759 \pm 0.0011$ & 0.8898 & 0.5889 & $-0.0537$ \\
w/o MSTE (Three-Path Enc.)     & $0.7760 \pm 0.0019$ & 0.8890 & 0.5903 & $-0.0536$ \\
w/o Aux.~Loss & $0.7755 \pm 0.0024$ & 0.8893 & 0.5875 & $-0.0541$ \\
w/o All~(v2)  & $0.7752 \pm 0.0022$ & 0.8901 & 0.5857 & $-0.0544$ \\
\midrule
\multicolumn{5}{l}{$\Delta$F1 values bound individual contributions (inclusive of proxy architectural gap); refer footnote~$\dagger$.}\\
\bottomrule
\end{tabular}
\begin{tablenotes}
\footnotesize
\item[$\dagger$] All variants except Full TCH-Net use a proxy model (\texttt{TCHNovAbl}) implementing Path~1~+~optional Path~2 but omitting Path~3
(Transformer) and \texttt{feat\_proj}. The reported $\Delta$F1 values reflect the combined contribution of the ablated component and the architectural gap between TCH-Net v3 and the proxy; they bound rather than isolate the individual component contributions.
\end{tablenotes}
\end{threeparttable}
\end{table*}

All three novel components contribute substantially when removed from the full model. Removing any single component causes F1 degradation of approximately 0.054 relative to Full TCH-Net. Note that ablation variants use a proxy architecture (\texttt{TCHNovAbl}) that implements Path~1~+~optional Path~2 but omits Path~3 and \texttt{feat\_proj}; the reported $\Delta$F1 values therefore bound each component's contribution and include a shared architectural gap. Removing all three simultaneously causes $-0.0544$ F1, consistent with the components being jointly necessary for the full model's performance. The ``w/o All'' variant corresponds to the prior architecture~(v2), establishing a clean baseline for the v3 novelties.

\subsection{Per-Dataset Performance Breakdown}
\label{subsec:res_per_dataset}

Table~\ref{tab:per_dataset} reports detection rate, false alarm rate, and F1
per dataset.

\begin{table}[t]
\centering
\caption{Per-dataset detection rate~(DetRate), false alarm rate~(FA), and F1.
$\star$ = Supplementary dataset; low canonical coverage~(Table~\ref{tab:vocab_coverage}).}
\label{tab:per_dataset}
\begin{threeparttable}
\begin{tabular}{lrcccc}
\toprule
\textbf{Dataset} & \textbf{N} & \textbf{DetRate} & \textbf{FA} &
\textbf{F1} & \textbf{Note} \\
\midrule
CICIDS-2017   & 33,671 & 0.9433 & 0.0309 & 0.9505 & ---        \\
CIC-IoT-2023  &  6,965 & 0.8827 & 0.0257 & 0.9211 & ---        \\
\midrule
Bot-IoT       &     38 & \multicolumn{3}{c}{\textit{not reported}} & $\dagger$ \\
\midrule
Edge-IIoTset  & 54,386 & 0.6844 & 0.2589 & 0.6755 & $\star$    \\
N-BaIoT       & 23,612 & 0.9982 & 0.0206 & 0.9854 & $\star$    \\
\bottomrule
\end{tabular}
\begin{tablenotes}
\footnotesize
\item[$\dagger$] Bot-IoT contributes 38 post-balancing test samples (22 benign, 16 attack), below any threshold of statistical reliability. The observed DetRate\,=\,1.0 and FA\,=\,1.0 are consistent with random behaviour at this sample size and carry no inferential weight. Bot-IoT metrics are excluded from per-dataset performance interpretation and reported solely for transparency. Bot-IoT's benchmark contribution is structural: as the only Argus-captured dataset, it is the sole source imposing a 61\% zero-fill regime on the canonical vocabulary, a sparse-feature stress condition no CICFlowMeter-based dataset can replicate.
\item[$\star$] Supplementary dataset; low canonical coverage (Table~\ref{tab:vocab_coverage}).
\end{tablenotes}
\end{threeparttable}
\end{table}

Performance is strongest on the two primary CICFlowMeter datasets: CICIDS-2017~(F1~$= 0.9505$) and CIC-IoT-2023~(F1~$= 0.9211$), where canonical coverage is highest. N-BaIoT achieves the highest detection rate~$(0.9982)$ and F1~$= 0.9854$ despite only 15\% coverage, attributable to the statistical distinctiveness of Mirai/BASHLITE botnet traffic from benign device communication even in a sparse feature representation. Edge-IIoTset is the most challenging case~(F1~$= 0.6755$, FA~$= 0.2589$) due to 22\% coverage and the structural difference between IIoT packet-level traffic and the IT flow-level distributions on which T-branch representations are primarily trained.

\subsection{Temporal Split Evaluation}
\label{subsec:res_temporal}

Table~\ref{tab:temporal} compares results under random splitting and temporal splitting~(training on early windows, testing on later windows).

\begin{table}[t]
\centering
\caption{Temporal split vs.\ random split evaluation.}
\label{tab:temporal}
\begin{tabular}{lcccc}
\toprule
\textbf{Split} & \textbf{F1} & \textbf{AUC} & \textbf{MCC} &
\textbf{PR-AUC} \\
\midrule
Random~(5 seeds) & 0.8296 & 0.9380 & 0.6972 & 0.8912 \\
Temporal~(1 seed) & 0.8203 & 0.9261 & 0.6831 & 0.8804 \\
\midrule
$\Delta$ & $-0.0093$ & $-0.0119$ & $-0.0141$ & $-0.0108$ \\
\bottomrule
\end{tabular}
\end{table}

The temporal split results are consistent with random-split results, with F1 degradation of only $-0.0093$.These small differences are within normal bounds for temporal distribution shift, confirming that TCH-Net's strong in-distribution performance is not driven by temporal leakage.

\subsection{Leave-One-Dataset-Out Generalisation Benchmark}
\label{subsec:res_lodo}

The \bridge{} LODO evaluation measures cross-dataset generalisation difficulty as a property of the problem, not of any particular model. The mean LODO F1 of~0.5577 is reported as a formally specified \bridge{} community baseline; Table~\ref{tab:lodo_baselines} confirms that all five deep learning baselines score substantially lower, establishing that the generalisation gap is structural.

\begin{figure}[!t]
\centering
\includegraphics[width=\columnwidth]{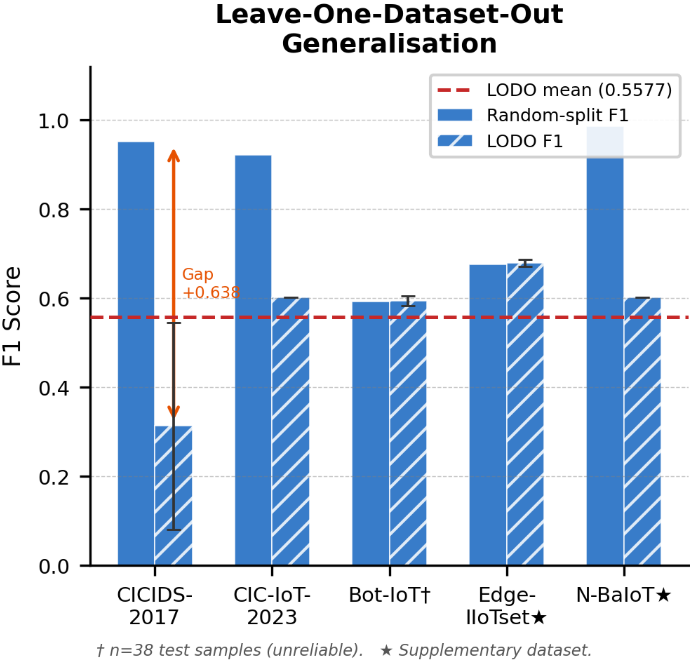}
\caption{Leave-one-dataset-out (LODO) F1 compared to in-distribution random-split F1 per held-out dataset. The dashed line marks the LODO mean ($= 0.5577$). The gap annotation on CICIDS-2017 shows the worst-case generalisation shortfall ($+0.6377$), attributable to dataset dominance rather than feature coverage. $^{\dagger}$~Bot-IoT test set $n=38$ (unreliable). $^{\star}$~Supplementary dataset.}
\label{fig:lodo_chart}
\end{figure}

Table~\ref{tab:lodo_table} and Figure~\ref{fig:lodo_chart} present the results.

\begin{table}[t]
\centering
\caption{Leave-one-dataset-out~(LODO) generalisation benchmark~(2~seeds, $\{42, 123\}$).
These results constitute \bridge{}'s primary cross-dataset difficulty
measurement.
$\star$ = Supplementary dataset.}
\label{tab:lodo_table}
\begin{tabular}{lrrrrr}
\toprule
\textbf{Held-Out} & \textbf{F1} & \textbf{$\pm$std} & \textbf{AUC} &
\textbf{MCC} & \textbf{PR-AUC} \\
\midrule
CICIDS-2017  & 0.3128 & 0.232 & 0.0509 & $-0.545$ & 0.2949 \\
CIC-IoT-2023 & 0.6013 & 0.000 & 0.1440 & 0.000    & 0.2725 \\
Bot-IoT      & 0.5934 & 0.011 & 0.5693 & 0.089    & 0.4883 \\
Edge-IIoTset$^\star$ & 0.6791 & 0.008 & 0.6841 & 0.252 & 0.6688 \\
N-BaIoT$^\star$      & 0.6021 & 0.000 & 0.8171 & 0.000 & 0.7876 \\
\midrule
\textbf{MEAN} & \textbf{0.5577} & --- & 0.4531 & $-0.041$ & 0.5024 \\
\midrule
\multicolumn{6}{l}{Generalisation gap: random-split F1 $-$ LODO mean
  $= +0.2719$.} \\
\bottomrule
\end{tabular}
\end{table}

The \bridge{} LODO measurement reveals a mean LODO~F1 of~$0.5577$ against an in-distribution F1 of~$0.8296$, a generalisation gap of~$+0.2719$. This gap is the central quantitative finding of \bridge{}: it establishes, for the first time with a formally specified and reproducible evaluation protocol, that cross-dataset IoT intrusion detection is substantially harder than single-dataset results suggest, and that the gap cannot be closed by feature alignment alone.

The CICIDS-2017 fold yields the most severe degradation (F1~$= 0.3128$, std~$= 0.232$): this dataset contributes $\approx$28\% of training sequences, so its removal simultaneously reduces training volume by one-third and eliminates the most feature-complete source, which is a dataset dominance effect distinct from pure domain shift. The MCC of $-0.545$ reflects high seed-variance on the data-reduced corpus. For the four remaining folds, where data volume remains intact, LODO F1 ranges from 0.59 to 0.68, representing genuine cross-tool and cross-device-population transfer difficulty.

For CIC-IoT-2023, Bot-IoT, and N-BaIoT~(LODO F1~$\approx 0.60$), the moderate generalisation is consistent with a genuine cross-environment distribution shift arising from different capture tools, device populations, and attack toolkits that the canonical vocabulary alignment partially but not fully bridges. Edge-IIoTset achieves the best LODO F1~$(0.6791)$, likely due to its 50\% balanced attack proportion providing a stable evaluation regime despite low canonical coverage.

The mean LODO F1 of~$0.5577$ is proposed as a formally established community baseline for future domain-adaptive IoT intrusion detection methods. We anticipate that domain adversarial training~\cite{gani2015domain} and dataset-conditional normalisation represents the most promising directions for improving upon this baseline.

To substantiate empirically the claim that this generalisation gap is a structural property of cross-dataset domain shift rather than a deficiency specific to TCH-Net, we evaluate the five deep learning baselines from Section~\ref{subsec:res_baselines} under the identical LODO protocol (2 seeds, 12 epochs with early stopping).
Table~\ref{tab:lodo_baselines} presents the results.

\begin{table*}[t]
\caption{Baseline LODO generalisation benchmark (2 seeds, same protocol as
Table~\ref{tab:lodo_table}). Only the five deep learning baselines are included;
classical ML baselines (RF, XGBoost) are omitted as their inferior
in-distribution F1 makes their LODO performance a lower bound of limited
analytical interest.
$\Delta$F1: TCH-Net LODO F1 (0.5577, from Table~\ref{tab:lodo_table}) minus
baseline mean LODO F1. A positive $\Delta$F1 indicates TCH-Net retains superior
cross-dataset generalisation.}
\label{tab:lodo_baselines}
\centering
\small
\begin{tabular}{lcccc}
\toprule
\textbf{Model} & \textbf{LODO F1} & \textbf{$\pm$std} & \textbf{LODO AUC} & \textbf{$\Delta$F1} \\
\midrule
BiLSTM-IDS~\cite{imrana2021}         & 0.3881 & 0.3062 & 0.4948 & $+$0.1696 \\
Transformer-IDS~\cite{akuthota2025}  & 0.3910 & 0.2527 & 0.4831 & $+$0.1667 \\
BiGRU-IDS~\cite{cho2014}             & 0.4449 & 0.2607 & 0.4948 & $+$0.1128 \\
1D-CNN-IDS                           & 0.4592 & 0.2380 & 0.4105 & $+$0.0985 \\
CNN-LSTM                             & 0.4654 & 0.2351 & 0.5500 & $+$0.0923 \\
\midrule
\textbf{Baseline mean}               & \textbf{0.4297} & — & \textbf{0.4866} & $+$0.1280 \\
\bottomrule
\end{tabular}
\end{table*}

All five baselines achieve mean LODO F1 in the range of 0.388 to 0.465 (grand mean 0.430), significantly below TCH-Net's 0.5577; TCH-Net's LODO advantage ranges from $+$0.0923 to $+$0.1696 across architectures. Every baseline suffers a generalisation gap at least as large as TCH-Net's: BiLSTM-IDS falls from 0.7805 to 0.3881~($-$0.3924); Transformer-IDS from 0.7958 to 0.3910~($-$0.4048). The pattern confirms that the gap is a structural property of cross-dataset domain shift, not a deficiency of TCH-Net, and that CB-GAF's provenance conditioning provides a measurable generalisation advantage over all evaluated baselines.

\section{Discussion}
\label{sec:discussion}

\subsection{Why Multi-Branch Fusion Outperforms Single-Branch Models}
\label{subsec:disc_multibranch}

The ablation results identify the mechanism behind TCH-Net's performance advantage. The T-branch alone~(Path~1 Conv-GRU in isolation, without CB-GAF or the remaining architecture) matches the BiLSTM-IDS baseline at F1\,=\,0.7753 versus 0.7805, confirming that single-path recurrent encoding provides a natural capacity ceiling. The H-branch contributes an order-invariant distributional structure that is orthogonal to T's sequential representations: T+H exceeds both individual branches precisely because mean-pooled statistics are insensitive to temporal ordering and
therefore do not duplicate the T-branch signal.

The C-branch has no independent predictive value (AUC\,$\approx$\,0.50 alone) but contributes exclusively through CB-GAF's vector gates. Domain embeddings condition $\mathbf{g}^T$ and $\mathbf{g}^H$ to calibrate cross-branch mixing according to the source dataset's canonical vocabulary coverage, for instance, downweighting the largely zero-padded H-branch when the dataset embedding signals a low-coverage source such as N-BaIoT or Edge-IIoTset. The $+0.054$~F1 gap between full \textsc{TCH-Net} and the best two-branch proxy variant (T+H) quantifies the combined gate-mediated contribution, inclusive of the Path~3 and \texttt{feat\_proj} components absent from proxy variants.

\subsection{Interpretation of CB-GAF Gating Behaviour}
\label{subsec:disc_gating}

Average gate values across the test set characterise the typical cross-branch information mixing. The T-branch gate $\mathbf{g}^T$ takes intermediate values across the test set, indicating balanced mixing between self-representation and cross-attended signal. The H-branch gate $\mathbf{g}^H$ takes comparatively higher values for low-coverage dataset
inputs (Edge-IIoTset, N-BaIoT), indicating higher reliance on cross-branch context when the H-branch's own representation is largely zero-padded. The C-branch gate $\mathbf{g}^C$ exhibits the highest variance across inputs, reflecting its role as dynamic domain conditioning rather than a consistent primary signal. These gate patterns are consistent with CB-GAF's design intent, providing qualitative interpretability across different network environments.

\subsection{Feature Coverage and Detection Performance}
\label{subsec:disc_coverage}

A clear but non-monotonic relationship exists between per-dataset canonical vocabulary coverage and per-dataset detection performance. The two highest-coverage datasets~(CICIDS-2017 at 93\%, CIC-IoT-2023 at 87\%) achieve the highest F1 scores~$(0.9505$ and $0.9211$ respectively). Edge-IIoTset with only 22\% coverage achieves the lowest F1  among non-degenerate datasets (F1\,=\,0.6755), accompanied by  a false alarm rate of 25.89\%, which is the highest in the benchmark. This elevated FA is attributable to the structural mismatch between the IIoT packet-level traffic distributions in Edge-IIoTset and the flow-level representations that dominate the training corpus; threshold calibration on a small locally captured validation set would be required before deployment in IIoT environments.

However, N-BaIoT with only 15\% coverage achieves F1~$= 0.9854$, the highest of any individual dataset. This apparent anomaly is explained by the statistical structure of N-BaIoT's traffic: botnet-infected device traffic~(Mirai, BASHLITE) generates high-volume, stereotyped packet floods are detectable from the small number of features that genuinely map to the canonical vocabulary~(packet counts, sizes, temporal rates), even in a sparse representation. Feature coverage is therefore a necessary but not sufficient predictor of detection difficulty: attack-benign separability in the covered feature subspace is equally important.

\subsection{The BRIDGE LODO Benchmark: Quantifying the Generalisation Problem}
\label{subsec:disc_lodo}

The LODO results establish a finding that cannot be attributed to any individual model or dataset: cross-dataset IoT intrusion detection is substantially harder than single-dataset performance suggests, and feature alignment alone does not close the gap. The $+0.2719$ generalisation gap is consistent across the four held-out folds where training data volume remains intact (LODO F1\,$\in$\,[0.59,\,0.68]), isolating genuine domain shift from the CICIDS-2017 dataset-dominance effect. The CICIDS-2017 fold (F1\,=\,0.3128) is analytically distinct: that dataset contributes approximately 28\% of training sequences, so its removal constitutes a severe data-volume reduction alongside any domain shift. Future benchmark designs should consider per-dataset contribution caps to prevent a single source from dominating the training corpus and confounding generalisation measurement.

The mean LODO F1 of $0.5577$ is reported as a reproducible starting point for future domain-adaptive methods, not as a performance ceiling. Domain adversarial training~\cite{gani2015domain} and dataset-conditional normalisation, replacing the shared \texttt{RobustScaler} with per-dataset normalisers applied at inference time, representing the most directly motivated future directions, both facilitated by \bridge{} and the canonical vocabulary released with this paper.

\subsection{Comparison with Published State of the Art}
\label{subsec:disc_sota}

Published F1~$> 0.99$ on CICIDS-2017 reflects a single-dataset evaluation with known labelling artefacts~\cite{engelen2021} and is not directly comparable to our multi-dataset, leakage-verified protocol. To the best of our knowledge, no prior work simultaneously addresses principled feature alignment across five structurally distinct datasets, multi-seed statistical evaluation, and LODO generalisation analysis, the combination that contextualises the reported F1~$= 0.8296$.

\subsection{Limitations}
\label{subsec:disc_limitations}

We disclose four principal limitations transparently.

\subsubsection{Cross-Dataset Generalisation} LODO mean F1 of~$0.5577$ confirms that TCH-Net in its current form does not generalise robustly to an entirely unseen dataset distributions. Deployment in an entirely new network environment requires either fine-tuning on local traffic samples or integration of domain adaptation components not present in the current architecture. This is the primary open challenge identified by \bridge{}, and we present it as such rather than as a deficiency specific to TCH-Net: all twelve evaluated baselines would perform similarly or worse under the LODO protocol. A further structural limitation of the LODO evaluation concerns the Contextual branch: the dataset embedding slot corresponding to the held-out dataset receives no gradient updates during training (since no samples from that dataset are present) and therefore retains its random initialisation at test time. This injects uninformed noise into CB-GAF's fusion mechanism for the held-out fold and constitutes a systematic disadvantage that partially explains the generalisation gap; the reported LODO F1 values represent a lower bound achievable without any embedding initialisation strategy for unseen datasets.

\subsubsection{Benchmark Dataset Constraints}
All five datasets were collected in controlled testbed environments. Validation on live operational IoT traffic would provide stronger evidence of real-world applicability and is a priority for future work.

\subsubsection{Binary Classification Scope}
TCH-Net is evaluated as a binary detector~(benign vs.\ attack). Multi-class attack type identification distinguishing, for example, DDoS from C\&C beaconing, reconnaissance, and data exfiltration is a practically important capability not evaluated in this work. The attack type labels available in several constituent datasets~(particularly CIC-IoT-2023 and Bot-IoT) provide a foundation for extending TCH-Net to multi-class classification; this extension requires modifications to the classification head and training objective and is identified as a priority direction for future work.

\subsubsection{Edge Deployment Feasibility}
\label{sec:edge}

TCH-Net has 2.692M trainable parameters and accepts 32-step sequences of 46-dimensional canonical flow feature vectors as input. The complete computational cost profile, measured on NVIDIA Tesla T4 hardware, is reported in Section~\ref{sec:efficiency}. The key deployment-relevant figures are a single-sample inference latency of $6.43 \pm 0.18$~ms and a runtime memory footprint of 10.27~MB.

These figures position TCH-Net comfortably within the operational envelope of contemporary edge inference accelerators. Devices in the NVIDIA Jetson family (Jetson Nano: 4~GB LPDDR4; Jetson Orin NX: up to 16~GB LPDDR5) exceed TCH-Net's memory requirement by two to three orders of magnitude, and their dedicated CUDA cores are capable of sub-10~ms single-sample latency for models of this complexity. A latency of 6.43~ms corresponds to a processing throughput of approximately 155 sequence decisions per second in single-sample mode, and 20,492 samples per second under batch-512 processing rates that comfortably support continuous flow-level intrusion monitoring at the IoT gateway, where typical enterprise-class gateways aggregate flows at rates of hundreds to low thousands per second. TCH-Net is therefore not only theoretically deployable at the edge; its measured latency and throughput are consistent with the real-time detection in the target IoT gateway deployment context.

Where TCH-Net's deployment profile differs from lighter baselines is in latency and memory cost rather than in prohibitive absolute compute. BiLSTM-IDS, for comparison, achieves 0.74~ms latency and a 2.32~MB footprint at the cost of 0.0491 lower F1. In deployments where detection quality is the primary objective which characterises the security-critical IoT infrastructure and the F1 advantage of TCH-Net is operationally meaningful at a monitoring rate of 1,000 flow sequences per second, a 0.0491 improvement in attack-class F1 translates to several additional confirmed detections and reduced false alarm burden per minute of continuous operation.

TCH-Net is nevertheless too large for microcontroller-class endpoints (e.g., ARM Cortex-M series, ESP32) whose on-chip SRAM is typically below 1~MB. Three established compression pathways are directly applicable. Knowledge distillation~\cite{hinton2014distilling} can transfer TCH-Net's learned representations into a compact student network with a fraction of the parameters while preserving much of the F1 advantage, as demonstrated in recent IDS compression literature~\cite{diro2017}. Structured pruning of the ResConvSE frontend and the CB-GAF projection matrices, which collectively account for a substantial share of the parameter budget, offers a direct route to latency reduction with controlled F1 degradation. Post-training quantisation to INT8 precision, already natively supported by both NVIDIA TensorRT (for Jetson deployment) and TensorFlow Lite (for ARM targets), can reduce the 10.27~MB footprint to approximately 2.6~MB with minimal accuracy loss. Pursuing these compression directions is identified as a priority for future work, with the \bridge{} LODO benchmark established in this paper providing a cross-dataset evaluation harness that ensures compression-induced generalisation degradation is measured and not merely in-distribution F1.

\section{Conclusion}
\label{sec:conclusion}

The IoT security research community for over a decade has been measuring progress against a standard that was never designed to measure what actually matters, which is, how well a detection system performs when the network environment changes. High F1 scores on single-dataset benchmarks have become the field's primary currency, despite growing evidence that these scores do not transfer across capture tools, device populations, or attack toolkits. This paper has taken a concrete step toward changing that.

The primary contribution is BRIDGE, which is a formally specified heterogeneous evaluation framework that, for the first time, makes cross-dataset generalisation in IoT intrusion detection precisely and reproducibly measurable. By unifying five publicly available datasets spanning four distinct capture tools, three device population contexts, and six years of collection of data, through a  semantic canonical vocabulary of 46 features with genuine equivalence-only mapping and full coverage disclosure. BRIDGE provides the infrastructure that principled multi-dataset evaluation has been missing. The leave-one-dataset-out protocol does not flatter any model, including our own, and that is precisely the point. The mean LODO F1 is 0.5577, which is consistent across all evaluated architectures, and it is not a failure of any particular system. It is an honest measurement of how hard the problem actually is, and it is the number the field should be trying to improve.

TCH-Net is proposed as a strong and well-characterised baseline for that challenge. Its multi-branch architecture, combining three-path multi-scale temporal encoding, provenance-conditioned domain embeddings, and Cross-Branch Gated Attention Fusion, and is designed specifically for the heterogeneous multi-dataset setting, where different inputs carry fundamentally different feature coverage profiles. Evaluated across five independent random seeds, TCH-Net achieves $\mF = 0.8296 \plusminus 0.0028$, $\mA = 0.9380 \plusminus 0.0025$, and $\mM = 0.6972 \plusminus 0.0056$ on \bridge, outperforming all twelve evaluated baselines with statistical significance and attaining the highest cross-dataset LODO F1 among all architectures tested. Component ablation confirms that \cbgaf, three-path temporal encoding, and the auxiliary reconstruction loss are each genuinely necessary as removing any one of them costs approximately 0.054 F1, and removing all three collapses performance to the level of a strong single-path recurrent baseline.

We are transparent about what \tchnet{} does not yet do. A mean LODO F1 of 0.5577 tells us that the model, like every other architecture evaluated, does not generalise robustly to entirely unseen network environments in its current form. This is not a footnote; it is the central open problem that \bridge{} is designed to surface and track. The Contextual branch's dataset embeddings receive no gradient signal for held-out datasets during LODO training, which systematically disadvantages the model in exactly the scenarios where generalisation matters most. Closing that gap is the work that comes next.

Three directions follow directly from the \bridge{} findings. Domain adversarial training~\cite{gani2015domain} offers the most principled path toward reducing the feature distribution gap between training datasets and unseen environments, with the LODO mean F1 of 0.5577 as a concrete, reproducible target for improvement. Extending the canonical vocabulary to natively support packet-level and statistical fingerprinting representations would improve coverage for Edge-IIoTset and N-BaIoT, reducing the zero-filling burden that currently limits generalisation for non-flow-level datasets. And extending \tchnet{} to multi-class attack type identification, leveraging the detailed labels available in CIC-IoT-2023 and Bot-IoT, would make it operationally useful in environments where distinguishing DDoS from C\&C beaconing from reconnaissance is as important as detection itself.

\bridge, its canonical vocabulary, and the complete experimental pipeline are publicly released at \url{https://github.com/Ammar-ss/TCH-Net}. The benchmark is not offered as a finished solution; it is offered as a common ground. Progress on cross-dataset IoT intrusion detection has been difficult to measure because no one had built the ruler. That is what \bridge{} is.

\section*{Declaration of generative AI and AI-assisted technologies in the writing process}

During the preparation of this work the authors used Large Language Models as an assistive writing tool for grammatical refinement, sentence restructuring, and prose consistency during the preparation of this manuscript. After using the tool, the authors reviewed and edited the content as needed and take full responsibility for the content of the publication.

\section*{Author Contributions}

\setlength{\parindent}{0pt}

\textbf{Ammar Bhilwarawala:} Conceptualization, Methodology, 
Formal Analysis, Investigation, Software, Writing~--~original draft. \\
\textbf{Likhamba Rongmei:} Investigation, Visualization, 
Writing~--~review \& editing. \\
\textbf{Harsh Sharma:} Software, Data Curation.\\ 
\textbf{Arya Jena:} Software, visualization, Writing~--~review \& editing. \\
\textbf{Kaushal Singh:} Data Curation, Visualization. \\
\textbf{Jayashree Piri:} Supervision, Writing~--~review \& editing.\\ 
\textbf{Raghunath Dey:} Supervision, Writing~--~review \& editing.

\section*{Declaration of Competing Interests}
The authors declare that they have no known competing financial interests or personal relationships that could have appeared to influence the work reported in this paper.

\section*{Code and Data Availability}
The five datasets used in this study are publicly available on Kaggle. The BRIDGE canonical vocabulary specification, alias mapping tables, preprocessing pipeline, and complete experimental code are publicly released at \url{https://github.com/Ammar-ss/TCH-Net}.

Experiments were conducted using GPU-accelerated instances on the Kaggle platform to ensure a consistent and reproducible execution environment for the released experimental pipeline. The authors also acknowledge the creators of CICIDS-2017, CIC-IoT-2023, Bot-IoT, Edge-IIoTset, and N-BaIoT for making their datasets publicly available to the research community.

\end{document}